
\input phyzzx
\endpage
\pagenumber=1
\input psfig
\def\havefigures{n}      
\catcode`\@=12        
%
\newbox\hdbox%
\newcount\hdrows%
\newcount\multispancount%
\newcount\ncase%
\newcount\ncols
\newcount\nrows%
\newcount\nspan%
\newcount\ntemp%
\newdimen\hdsize%
\newdimen\newhdsize%
\newdimen\parasize%
\newdimen\spreadwidth%
\newdimen\thicksize%
\newdimen\thicksz
\newdimen\thinsize%
\newdimen\tablewidth%
\newif\ifcentertables%
\newif\ifendsize%
\newif\iffirstrow%
\newif\iftableinfo%
\newtoks\dbt%
\newtoks\hdtks%
\newtoks\savetks%
\newtoks\tableLETtokens%
\newtoks\tabletokens%
\newtoks\widthspec%
%
%
%
%
\tableinfotrue%
\catcode`\@=11
%
%
\def\tstrut{\vrule height3.1ex depth1.2ex width0pt}%
\def\and{\char`\&}
\def\tablerule{\noalign{\hrule height\thinsize depth0pt}}%
\thicksize=1.5pt
\thinsize=0.6pt
\thicksz=1.5pt
\def\thickrule{\noalign{\hrule height\thicksize depth0pt}}%
\def\ctr#1{\hfil\ #1\hfil}%
%
%
%
%
\tablewidth=-\maxdimen%
\spreadwidth=-\maxdimen%
\def\tabskipglue{0pt plus 1fil minus 1fil}%
%
%
\centertablestrue%
%
%
%
%
\parasize=4in%
\gdef\ARGS{########}
\gdef\headerARGS{####}
\def\@mpersand{&}
{\catcode`\|=13
\gdef\letbarzero{\let|0}
\gdef\letbartab{\def|{&&}}%
\gdef\letvbbar{\let\vb|}%
}
{\catcode`\&=4
\def\ampskip{&\omit\hfil&}
\catcode`\&=13
\let&0
\xdef\letampskip{\def&{\ampskip}}%
\gdef\letnovbamp{\let\novb&\let\tab&}
}
\def\begintable{
   \begingroup%
   \catcode`\|=13\letbartab\letvbbar%
   \catcode`\&=13\letampskip\letnovbamp%
   \def\multispan##1{
      \omit \mscount##1%
      \multiply\mscount\tw@\advance\mscount\m@ne%
      \loop\ifnum\mscount>\@ne \sp@n\repeat%
   }
   \def\|{%
      &\omit\widevline&%
   }%
   \ruledtable
}
\long\def\ruledtable#1\endtable{%
%
%
%
   \offinterlineskip
   \tabskip 0pt
   \def\widevline{\vrule width\thicksz}
   \def\endrow{\@mpersand\omit\hfil\crnorm\@mpersand}%
   \def\crthick{\@mpersand\crnorm\thickrule\@mpersand}%
   \def\crthickneg##1{\@mpersand\crnorm\thickrule
	  \noalign{{\skip0=##1\vskip-\skip0}}\@mpersand}%
   \def\crnorule{\@mpersand\crnorm\@mpersand}%
   \def\crnoruleneg##1{\@mpersand\crnorm
	  \noalign{{\skip0=##1\vskip-\skip0}}\@mpersand}%
   \let\nr=\crnorule
   \def\endtable{\@mpersand\crnorm\thickrule}%
   \let\crnorm=\cr
%
%
   \edef\cr{\@mpersand\crnorm\tablerule\@mpersand}%
   \def\crneg##1{\@mpersand\crnorm\tablerule
	  \noalign{{\skip0=##1\vskip-\skip0}}\@mpersand}%
   \let\ctneg=\crthickneg
   \let\nrneg=\crnoruleneg
   \the\tableLETtokens
%
%
   \tabletokens={&#1}
%
%
   \countROWS\tabletokens\into\nrows%
   \countCOLS\tabletokens\into\ncols%
%
%
   \advance\ncols by -1%
   \divide\ncols by 2%
   \advance\nrows by 1%
%
%
   \iftableinfo %
      \immediate\write16{[Nrows=\the\nrows, Ncols=\the\ncols]}%
   \fi%
%
%
   \ifcentertables
      \ifhmode \par\fi
      \line{
      \hss
   \else %
      \hbox{%
   \fi
      \vbox{%
	 \makePREAMBLE{\the\ncols}
	 \edef\next{\preamble}
	 \let\preamble=\next
	 \makeTABLE{\preamble}{\tabletokens}
      }
      \ifcentertables \hss}\else }\fi
   \endgroup
   \tablewidth=-\maxdimen
   \spreadwidth=-\maxdimen
}
\def\makeTABLE#1#2{
   {
   \let\ifmath0
   \let\header0
   \let\multispan0
%
%
   \ncase=0%
   \ifdim\tablewidth>-\maxdimen \ncase=1\fi%
   \ifdim\spreadwidth>-\maxdimen \ncase=2\fi%
   \relax
%
   \ifcase\ncase %
      \widthspec={}%
   \or %
      \widthspec=\expandafter{\expandafter t\expandafter o%
		 \the\tablewidth}%
   \else %
      \widthspec=\expandafter{\expandafter s\expandafter p\expandafter r%
		 \expandafter e\expandafter a\expandafter d%
		 \the\spreadwidth}%
   \fi %
   \xdef\next{
      \halign\the\widthspec{%
      #1
      \noalign{\hrule height\thicksize depth0pt}
%
      \the#2\endtable
%
      }
   }
   }
   \next
}
\def\makePREAMBLE#1{
   \ncols=#1
   \begingroup
   \let\ARGS=0
   \edef\xtp{\widevline\ARGS\tabskip\tabskipglue%
   &\ctr{\ARGS}\tstrut}
   \advance\ncols by -1
   \loop
      \ifnum\ncols>0 %
      \advance\ncols by -1%
      \edef\xtp{\xtp&\vrule width\thinsize\ARGS&\ctr{\ARGS}}%
   \repeat
   \xdef\preamble{\xtp&\widevline\ARGS\tabskip0pt%
   \crnorm}
   \endgroup
}
\def\countROWS#1\into#2{
   \let\countREGISTER=#2%
   \countREGISTER=0%
   \expandafter\ROWcount\the#1\endcount%
}%
\def\ROWcount{%
   \afterassignment\subROWcount\let\next= %
}%
\def\subROWcount{%
   \ifx\next\endcount %
      \let\next=\relax%
   \else%
      \ncase=0%
      \ifx\next\cr %
	 \global\advance\countREGISTER by 1%
	 \ncase=0%
      \fi%
      \ifx\next\endrow %
	 \global\advance\countREGISTER by 1%
	 \ncase=0%
      \fi%
      \ifx\next\crthick %
	 \global\advance\countREGISTER by 1%
	 \ncase=0%
      \fi%
      \ifx\next\crnorule %
	 \global\advance\countREGISTER by 1%
	 \ncase=0%
      \fi%
      \ifx\next\crthickneg %
	 \global\advance\countREGISTER by 1%
	 \ncase=0%
      \fi%
      \ifx\next\crnoruleneg %
	 \global\advance\countREGISTER by 1%
	 \ncase=0%
      \fi%
      \ifx\next\crneg %
	 \global\advance\countREGISTER by 1%
	 \ncase=0%
      \fi%
      \ifx\next\header %
	 \ncase=1%
      \fi%
      \relax%
      \ifcase\ncase %
	 \let\next\ROWcount%
      \or %
	 \let\next\argROWskip%
      \else %
      \fi%
   \fi%
   \next%
}
\def\counthdROWS#1\into#2{%
\dvr{10}%
   \let\countREGISTER=#2%
   \countREGISTER=0%
\dvr{11}%
\dvr{13}%
   \expandafter\hdROWcount\the#1\endcount%
\dvr{12}%
}%
\def\hdROWcount{%
   \afterassignment\subhdROWcount\let\next= %
}%
\def\subhdROWcount{%
   \ifx\next\endcount %
      \let\next=\relax%
   \else%
      \ncase=0%
      \ifx\next\cr %
	 \global\advance\countREGISTER by 1%
	 \ncase=0%
      \fi%
      \ifx\next\endrow %
	 \global\advance\countREGISTER by 1%
	 \ncase=0%
      \fi%
      \ifx\next\crthick %
	 \global\advance\countREGISTER by 1%
	 \ncase=0%
      \fi%
      \ifx\next\crnorule %
	 \global\advance\countREGISTER by 1%
	 \ncase=0%
      \fi%
      \ifx\next\header %
	 \ncase=1%
      \fi%
\relax%
      \ifcase\ncase %
	 \let\next\hdROWcount%
      \or%
	 \let\next\arghdROWskip%
      \else %
      \fi%
   \fi%
   \next%
}%
{\catcode`\|=13\letbartab
\gdef\countCOLS#1\into#2{%
   \let\countREGISTER=#2%
   \global\countREGISTER=0%
   \global\multispancount=0%
   \global\firstrowtrue
   \expandafter\COLcount\the#1\endcount%
   \global\advance\countREGISTER by 3%
   \global\advance\countREGISTER by -\multispancount
}%
\gdef\COLcount{%
   \afterassignment\subCOLcount\let\next= %
}%
{\catcode`\&=13%
\gdef\subCOLcount{%
   \ifx\next\endcount %
      \let\next=\relax%
   \else%
      \ncase=0%
      \iffirstrow
	 \ifx\next& %
	    \global\advance\countREGISTER by 2%
	    \ncase=0%
	 \fi%
	 \ifx\next\span %
	    \global\advance\countREGISTER by 1%
	    \ncase=0%
	 \fi%
	 \ifx\next| %
	    \global\advance\countREGISTER by 2%
	    \ncase=0%
	 \fi
	 \ifx\next\|
	    \global\advance\countREGISTER by 2%
	    \ncase=0%
	 \fi
	 \ifx\next\multispan
	    \ncase=1%
	    \global\advance\multispancount by 1%
	 \fi
	 \ifx\next\header
	    \ncase=2%
	 \fi
	 \ifx\next\cr	    \global\firstrowfalse \fi
	 \ifx\next\endrow   \global\firstrowfalse \fi
	 \ifx\next\crthick  \global\firstrowfalse \fi
	 \ifx\next\crnorule \global\firstrowfalse \fi
	 \ifx\next\crnoruleneg \global\firstrowfalse \fi
	 \ifx\next\crthickneg  \global\firstrowfalse \fi
	 \ifx\next\crneg       \global\firstrowfalse \fi
      \fi
\relax
      \ifcase\ncase %
	 \let\next\COLcount%
      \or %
	 \let\next\spancount%
      \or %
	 \let\next\argCOLskip%
      \else %
      \fi %
   \fi%
   \next%
}%
\gdef\argROWskip#1{%
   \let\next\ROWcount \next%
}
\gdef\arghdROWskip#1{%
   \let\next\ROWcount \next%
}
\gdef\argCOLskip#1{%
   \let\next\COLcount \next%
}
}
}
\def\spancount#1{
   \nspan=#1\multiply\nspan by 2\advance\nspan by -1%
   \global\advance \countREGISTER by \nspan
   \let\next\COLcount \next}%
\def\dvr#1{\relax}%
\def\header#1{%
\dvr{1}{\let\cr=\@mpersand%
\hdtks={#1}%
\counthdROWS\hdtks\into\hdrows%
\advance\hdrows by 1%
\ifnum\hdrows=0 \hdrows=1 \fi%
\dvr{5}\makehdPREAMBLE{\the\hdrows}%
\dvr{6}\getHDdimen{#1}%
{\parindent=0pt\hsize=\hdsize{\let\ifmath0%
\xdef\next{\valign{\headerpreamble #1\crnorm}}}\dvr{7}\next\dvr{8}%
}%
}\dvr{2}}
\def\makehdPREAMBLE#1{
\dvr{3}%
\hdrows=#1
{
\let\headerARGS=0%
\let\cr=\crnorm%
\edef\xtp{\vfil\hfil\hbox{\headerARGS}\hfil\vfil}%
\advance\hdrows by -1
\loop
\ifnum\hdrows>0%
\advance\hdrows by -1%
\edef\xtp{\xtp&\vfil\hfil\hbox{\headerARGS}\hfil\vfil}%
\repeat%
\xdef\headerpreamble{\xtp\crcr}%
}
\dvr{4}}
\def\getHDdimen#1{%
\hdsize=0pt%
\getsize#1\cr\end\cr%
}
\def\getsize#1\cr{%
\endsizefalse\savetks={#1}%
\expandafter\lookend\the\savetks\cr%
\relax \ifendsize \let\next\relax \else%
\setbox\hdbox=\hbox{#1}\newhdsize=1.0\wd\hdbox%
\ifdim\newhdsize>\hdsize \hdsize=\newhdsize \fi%
\let\next\getsize \fi%
\next%
}%
\def\lookend{\afterassignment\sublookend\let\looknext= }%
\def\sublookend{\relax%
\ifx\looknext\cr %
\let\looknext\relax \else %
   \relax
   \ifx\looknext\end \global\endsizetrue \fi%
   \let\looknext=\lookend%
    \fi \looknext%
}%
%
%
\def\tablelet#1{%
   \tableLETtokens=\expandafter{\the\tableLETtokens #1}%
}%
\catcode`\@=12
%


    \font\eighti=cmmi8                          \skewchar\eighti='177
    \font\eightsy=cmsy8                          \skewchar\eightsy='60
    \font\eightsl=cmsl8
    \font\eightit=cmti8
    \def\noblackbox{\overfullrule=0pt}
    \noblackbox
    \def\half{{1\over2}}

    \def\bold#1{\setbox0=\hbox{$#1$}%
         \kern-.025em\copy0\kern-\wd0
         \kern.05em\copy0\kern-\wd0
         \kern-.025em\raise.0433em\box0 }
    \def\unlock{\catcode`@=11} 
    \def\lock{\catcode`@=12} 
    \def\Buildrel#1\under#2{\mathrel{\mathop{#2}\limits_{#1}}}
    \def\llongrarrow{\hbox to 40pt{\rightarrowfill}}

    %
     \newtoks\slashfraction
     \slashfraction={.13}
     \def\slash#1{\setbox0\hbox{$ #1 $}
     \setbox0\hbox to \the\slashfraction\wd0{\hss \box0}/\box0 }
     \unlock
     \def\leftrightarrowfill{$\m@th\mathord-\mkern-6mu%
       \cleaders\hbox{$\mkern-2mu\mathord-\mkern-2mu$}\hfill
       \mkern-6mu\mathord\leftrightarrow$}
     \def\overlrarrow#1{\vbox{\ialign{##\crcr
           \leftrightarrowfill\crcr\noalign{\kern-\p@\nointerlineskip}
           $\hfil\displaystyle{#1}\hfil$\crcr}}}
     \lock
    %

    {\obeyspaces\global\let =\ }   
    %
    \def\papersize{       \hsize=35pc\vsize=50pc\hoffset=1cm\voffset=1.3cm
                  \pagebottomfiller=0pc
                  \skip\footins=\bigskipamount\normalspace}
    \def\lettersize{\hsize=6.5in\vsize=8.5in\hoffset=0cm\voffset=1.6cm
                  \pagebottomfiller=\letterbottomskip
                  \skip\footins=\smallskipamount
                  \multiply\skip\footins by 3
                  \singlespace}
    \papers
    %
    \catcode`\@=11 
    \newif\ifletterstyle                
    \letterstylefalse             
    \def\letters{\lettersize\letterstyletrue
       \headline=\letterheadline \footline=\letterfootline
       \immediate\openout\labelswrite=\jobname.lab}
    \def\iftpub{\afterassignment\iftp@b\toks@}
    \def\iftp@b{\edef\n@xt{\Pubnum={UFIFT-HEP--\the\toks@}}\n@xt}
    \let\pubnum=\iftpub
    \expandafter\ifx\csname eightrm\endcsname\relax
        \let\eightrm=\ninerm \let\eightbf=\ninebf \fi
    \catcode`\@=12 
    %
       
     

    \unlock
    \def\eightpoint{\relax
        \textfont0=\eightrm          \scriptfont0=\eightrm
        \scriptscriptfont0=\fiverm
        \def\rm{\fam0 \eightrm \f@ntkey=0 }\relax
        \textfont1=\eighti           \scriptfont1=\eighti
        \scriptscriptfont1=\fivei
        \def\oldstyle{\fam1 \eighti \f@ntkey=1 }\relax
        \textfont2=\eightsy          \scriptfont2=\eightsy
        \scriptscriptfont2=\fivesy
        \textfont3=\tenex          \scriptfont3=\tenex
        \scriptscriptfont3=\tenex
        \def\it{\fam\itfam \eightit \f@ntkey=4 }\textfont\itfam=\eightit
        \def\sl{\fam\slfam \eightsl \f@ntkey=5 }\textfont\slfam=\eightsl
        \def\bf{\fam\bffam \eightbf \f@ntkey=6 }\textfont\bffam=\eightbf
            \scriptfont\bffam=\eightbf     \scriptscriptfont\bffam=\fivebf
        \def\tt{\fam\ttfam \eighttt \f@ntkey=7 }\textfont\ttfam=\eighttt
        \setbox\strutbox=\hbox{\vrule height 4pt depth 3pt width\z@}
        \samef@nt}
    \lock
    \def\boxit#1{\vbox{\hrule\hbox{\vrule\kern3pt
                 \vbox{\kern3pt#1\kern3pt}\kern3pt\vrule}\hrule}}
     \newdimen\str

    \def\fboxit#1#2{\vbox{\hrule height #1 \hbox{\vrule width #1
               \kern3pt \vbox{\kern3pt#2\kern3pt}\kern3pt \vrule width #1 }
               \hrule height #1 }}
    
    \def\fillbox#1{\hbox to #1{\vbox to #1{\vfil}\hfil}}
    \def\dotbox#1{\hbox to #1{\vbox to 10pt{\vfil}\hss $\cdots$ \hss}}
    \def\ggenbox#1#2{\vbox to 10pt{\vss \hbox to #1{\hss #2  \hss} \vss}}

    
    \catcode`\@=11 
    \newtoks\foottokens
    \let\labelfont=\Tenpoint      
    \def\MakeFromBox{\gl@bal\setbox\FromLabelBox=\vbox{\labelfont
         \ialign{##\hfil\cr \the\sendername \the\FromAddress \crcr }}}
    \def\smallsize{\relax
    \def\eightpoint{\relax
    \textfont0=\eightrm  \scriptfont0=\sixrm
    \scriptscriptfont0=\fiverm
    \def\rm{\fam0 \eightrm \f@ntkey=0}\relax
    \textfont1=\eighti  \scriptfont1=\sixi
    \scriptscriptfont1=\fivei
    \def\oldstyle{\fam1 \eighti \f@ntkey=1}\relax
    \textfont2=\eightsy  \scriptfont2=\sixsy
    \scriptscriptfont2=\fivesy
    \textfont3=\tenex  \scriptfont3=\tenex
    \scriptscriptfont3=\tenex
        \def\it{\fam\itfam \eightit \f@ntkey=4 }\textfont\itfam=\eightit
    \def\sl{\fam\slfam \eightsl \f@ntkey=5 }\textfont\slfam=\eightsl
    \def\bf{\fam\bffam \eightbf \f@ntkey=6 }\textfont\bffam=\eightbf
    \scriptfont\bffam=\sixbf   \scriptscriptfont\bffam=\sixbf
    \def\tt{\fam\ttfam \eighttt \f@ntkey=7 }
    \def\caps{\fam\cpfam \tencp \f@ntkey=8 }\textfont\cpfam=\tencp
    \setbox\strutbox=\hbox{\vrule height 7.35pt depth 3.02pt width\z@}
    \samef@nt}
    \normalbaselineskip = 16.60pt plus 0.166pt minus 0.083pt
    \normallineskip = 1.25pt plus 0.08pt minus 0.08pt
    \normallineskiplimit = 1.25pt
    \normaldisplayskip = 16.60pt plus 4.15pt minus 8.3pt
    \normaldispshortskip = 4.98pt plus 3.32pt
    \normalparskip = 4.98pt plus 1.67pt minus .83pt
    \skipregister = 4.15pt plus 1.67pt minus 1.25pt
    \def\Eightpoint{\eightpoint \relax
      \ifsingl@\subspaces@t2:5;\else\subspaces@t3:5;\fi
      \ifdoubl@ \multiply\baselineskip by 5
                \divide\baselineskip by 4\fi }
    \parindent=16.67pt
    \itemsize=25pt
    \thinmuskip=2.5mu
    \medmuskip=3.33mu plus 1.67mu minus 3.33mu
        \thickmuskip=4.17mu plus 4.17mu
    \def\thinspace{\kern .13889em }
    \def\negthinspace{\kern-.13889em }
    \def\enspace{\kern.416667em }
    \def\enskip{\hskip.416667em\relax}
    \def\quad{\hskip.83333em\relax}
    \def\qquad{\hskip1.66667em\relax}
    \def\crr{\cropen{8.3333pt}}
    \labelwidth=4.5in
    \let\labelfont=\Eightpoint
    \let\letterhead=\FLOHEAD      
    \def\Vfootnote##1{\insert\footins\bgroup
       \interlinepenalty=\interfootnotelinepenalty \floatingpenalty=20000
       \singl@true\doubl@false\Eightpoint
       \splittopskip=\ht\strutbox \boxmaxdepth=\dp\strutbox
       \leftskip=\footindent \rightskip=\z@skip
       \parindent=0.5\footindent \parfillskip=0pt plus 1fil
       \spaceskip=\z@skip \xspaceskip=\z@skip \footnotespecial
       \Textindent{##1}\footstrut\futurelet\next\fo@t}%
    \def\attach##1{\step@ver{\strut^{\mkern 1.6667mu ##1} } }
    \def\inserttable ##1##2##3%
        {%
        \tbldef {##1}{##3}\goodbreak%
            \midinsert
          \smallskip
          \hbox{\singlespace \hskip 0.5cm
                  \vtop{\parshape=2 0cm 10.8cm 1.3cm 9.5cm
                        \noindent{\bf\Table{##1}}.\enspace ##3}
                  \hfil}
          ##2
          \smallskip
        \endinsert
        }
    \def\sure{y}
    \def\insertfigure ##1##2##3%
        {%
        \figdef {##1}{##3}\goodbreak%
        \midinsert
          \smallskip
          ##2
          \hbox{\singlespace\hskip 0.5cm
                  \vtop{\parshape=2 0cm 10.8cm
                          1.6cm 9.2cm \noindent{\bf\Figure{##1}}.
                          \enspace ##3}
                  \hfil}
            \smallskip
        \endinsert
        }%
    \def\references{\par\penalty-300\vskip\chapterskip
            \spacecheck\chapterminspace
          \line{\twelverm\hfil REFERENCES\hfil}
          \nobreak\vskip\headskip\penalty 30000
          \reflist{}}
    \def\figures{\par\penalty-300\vskip\chapterskip
            \spacecheck\chapterminspace
          \line{\twelverm\hfil FIGURE CAPTIONS\hfil}
          \nobreak\vskip\headskip\penalty 30000
          \figlist{}}
    \def\tables{\par\penalty-300\vskip\chapterskip
            \spacecheck\chapterminspace
          \line{\twelverm\hfil TABLE CAPTIONS\hfil}
          \nobreak\vskip\headskip\penalty 30000
          \tbllist{}}
    \def\PH@SR@V{\doubl@true\baselineskip=20.08pt plus .1667pt minus .0833pt
                 \parskip = 2.5pt plus 1.6667pt minus .8333pt }
        \def\author##1{\vskip\frontpageskip\titlestyle{\tencp ##1}\nobreak}
    \def\address##1{\par\kern 4.16667pt\titlestyle{\tenpoint\it ##1}}
    \def\andaddress{\par\kern 4.16667pt \centerline{\sl and} \address}
    \def\UFL{\address{Department of Physics\break
          University of Florida, Gainesville, FL 32611}}
    \def\abstract{\vskip\frontpageskip\centerline{\twelverm ABSTRACT}
                  \vskip\headskip }
    \def\submit##1{\par\nobreak\vfil\nobreak\medskip
       \centerline{Submitted to \sl ##1}}
    \def\doeack{\foot{Work supported by the Department of Energy,
                  contract  DE--FG05--86ER--40272.}}
    \def\nsfack{\foot{Work supported by National Science Foundation
                  Grant  PHY 84--16030A01.}}
    \def\cases##1{\left\{\,\vcenter{\Tenpoint\m@th
        \ialign{$####\hfil$&\quad####\hfil\crcr##1\crcr}}\right.}
    \def\matrix##1{\,\vcenter{\Tenpoint\m@th
        \ialign{\hfil$####$\hfil&&\quad\hfil$####$\hfil\crcr
          \mathstrut\crcr\noalign{\kern-\baselineskip}
         ##1\crcr\mathstrut\crcr\noalign{\kern-\baselineskip}}}\,}
    \Tenpoint
        }
    \newdimen\fullhsize
    \newbox\leftcolumn
    \def\twoinone{
    \smallsize
    \def\papersize{
                  \voffset=-.23truein
                  \vsize=7truein
                  \baselineskip=16pt plus 2pt minus 1pt
                  \fullhsize=10truein\hsize=4.75truein
                    \hoffset=-.54truein
                  \skip\footins=\bigskipamount}
    \def\lettersize{\voffset=.31truein
                  \vsize=6.38truein
                  \baselineskip=16pt plus 2pt minus 1pt
                  \fullhsize=10truein\hsize=4.75truein
                    \hoffset=-.48truein
                  \skip\footins=\smallskipamount
                    \multiply\skip\footins by3}
    \papers               
        \let\lr=L
    \output={\if L\lr
                  \global\setbox\leftcolumn=\columnbox \advancepageno
                  \global\let\lr=R
           \else  \getitout \advancepageno
                  \global\let\lr=L\fi
           \ifnum\outputpenalty>-20000 \else\dosupereject\fi}
    }             
          \def\columnbox{\leftline{
                  \vbox{\ifletterstyle\makeheadline\fi
                          \pagebody\makefootline}}}
          \def\fullline{\hbox to\fullhsize}
          \def\getitout{\shipout\vbox{\fullline{\box\leftcolumn
                  \hfil {\leftline{
                  \vbox{\makeheadline
                  \pagebody\makefootline}}} }}}
    %
    \catcode`\@=12 
        %
    %
    %
    \newcount      \ObjClass
    \chardef\ClassNum     = 0
    \chardef\ClassMisc    = 1
    \chardef\ClassEqn     = 2
    \chardef\ClassRef     = 3
    \chardef\ClassFig     = 4
    \chardef\ClassTbl     = 5
    \chardef\ClassThm     = 6
    \chardef\ClassStyle     = 7
        \chardef\ClassDef       = 8
    \edef\NumObj  {\ObjClass = \ClassNum   \relax}
    \edef\MiscObj {\ObjClass = \ClassMisc  \relax}
    \edef\EqnObj  {\ObjClass = \ClassEqn   \relax}
    \edef\RefObj  {\ObjClass = \ClassRef   \relax}
    \edef\FigObj  {\ObjClass = \ClassFig   \relax}
    \edef\TblObj  {\ObjClass = \ClassTbl   \relax}
    
    \edef\StyleObj  {\ObjClass = \ClassStyle \relax}
    \edef\DefObj    {\ObjClass = \ClassDef   \relax}
    %
    %
    \def\gobble    #1{}%
    \def\trimspace   #1 \end{#1}%
    \def\ifundefined #1{\expandafter \ifx \csname#1\endcsname \relax}%
    \def\trimprefix  #1_#2\end{\expandafter \string \csname #2\endcsname}%
    \def\skipspace #1#2#3\end%
        {%
        \def \temp {#2}%
        \ifx \temp\space \skipspace #1#3\end
        \else \gdef #1{#2#3}\fi
            }%
    \def\stylename#1{\expandafter\expandafter\expandafter
        \gobble\expandafter\string\the#1}
    \ifundefined {protect} \let\protect=\relax \fi
    \catcode`\@=11
    \let\rel@x=\relax
    \def\relaxtest{\rel@x}
    \catcode`\@=12
    \def\checkchapterlabel%
        {%
            {\protect\if\chapterlabel\relaxtest
          \global\let\chapterlabel=\relax\fi}
        }%
    \begingroup
    \catcode`\<=1 \catcode`\{=12
    \catcode`\>=2 \catcode`\}=12
    \xdef\LBrace<{>%
    \xdef\RBrace<}>%
    \endgroup
    %
    %
    \newcount\equanumber \equanumber=0
        \newcount\eqnumber   \eqnumber=0
    \newif\ifleftnumbers \leftnumbersfalse

    \def\(#1)%
         {%
            \ifnum \equanumber<0 \eqnumber=-\equanumber
              \advance\eqnumber by -1 \else
                \eqnumber = \equanumber\fi
            \ifmmode\ifinner(\eqnum{#1})\else
            \ifleftnumbers\leqno(\eqnum{#1})\ifdraft{\rm[#1]}\fi
                \else\eqno(\eqnum{#1})\ifdraft{\rm[#1]}\fi\fi\fi
          \else(\eqnum{#1})\fi\ifnum%
              \equanumber<0 \global\equanumber=-\eqnumber\global\advance
                \equanumber by -1\else\global\equanumber=\eqnumber\fi
         }%
    \def\mideq(#1)%
         {%
          \ifleftnumbers \leqinsert{$\(#1)$} \else
          \eqinsert{$\(#1)$} \fi
         }%
    \def\eqnum #1%
        {%
        \LookUp Eq_#1 \using\eqnumber\neweqnum
        {\rm\label}%
        }%
    \def\neweqnum #1#2%
        {%
        \checkchapterlabel
        {\protect\xdef\eqnoprefix{\ifundefined{chapterlabel}
          \else\chapterlabel.\fi}}
        \ifmmode \xdef #1{\eqnoprefix #1}
            \else\message{Undefined equation \string#1 in non-math mode.}%
               \xdef #1{\relax}
               \global\advance \eqnumber by -1
            \fi
        \EqnObj \SaveObject{#1}{#2}
        }%
    \everydisplay = {\expandafter \let\csname Eq_\endcsname=\relax
                   \expandafter \let\csname Eq_?\endcsname=\relax}%
    %
        %
    \newcount\tablecount \tablecount=0
    \def\Table  #1{Table~\tblnum {#1}}%
    \def\tblnum #1{\TblObj \LookUp Tbl_#1 \using\tablecount
          \SaveObject \label\ifdraft [#1]\fi}%
    \def\tbldef #1{\TblObj \SaveContents {Tbl_#1}}%
    \def\tbllist  {\TblObj \ListObjects}%
    %
    \def\inserttable #1#2#3%
        {%
        \tbldef {#1}{#3}\goodbreak%
        \midinsert
          \smallskip
          \hbox{\singlespace
                \vtop{\titlestyle{{\Tenpoint{\caps\Table{#1}}\break #3}}}
               }%
          #2
          \smallskip%
        \endinsert
        }%
        \def\topinserttable #1#2#3%
        {%
        \tbldef {#1}{#3}\goodbreak%
        \topinsert
          \smallskip
          \hbox{\singlespace
                \vtop{\titlestyle{{\Tenpoint{\caps\Table{#1}}\break #3}}}
               }%
          #2
          \smallskip%
        \endinsert
        }%
    %
    %
    \newcount\figurecount \figurecount=0
    \def\Figure #1{Figure~\fignum {#1}}%
    \def\Fig    #1{Fig.~\fignum {#1}}%
    \def\fignum #1{\FigObj \LookUp Fig_#1 \using\figurecount
         \SaveObject \label\ifdraft [#1]\fi}%
        \def\figdef #1{\FigObj \SaveContents {Fig_#1}}%
    \def\figlist  {\FigObj \ListObjects}%
    %
    \def\sure{y}
    \def\insertfigure #1#2#3#4%
        {%
        \figdef {#1}{#3}%
        \midinsert
          \bigskip
          \ifx\havefigures\sure
          #2
          \else
             {#4}\fi
          \hbox{  \singlespace
                  \hskip 0.4in
                  \vtop{\parshape=2 0pt 362pt 32pt 330pt
                        \noindent{\Tenpoint{\caps\Fig{#1}}.\enspace #3}}
                  \hfil}
          \smallskip%
        \endinsert
        }%
    \def\topinsertfigure #1#2#3%
        {%
            \figdef {#1}{#3}%
        \topinsert
          \bigskip
          #2
          \hbox{  \singlespace
                  \hskip 0.4in
                  \vtop{\parshape=2 0pt 362pt 32pt 330pt
                        \noindent{\Tenpoint{\caps\Fig{#1}}.\enspace #3}}
                  \hfil}
          \smallskip%
        \endinsert
        }%
    %
    %
    %
        \newcount\theoremcount \theoremcount=0
    %
    %
    %
    %
    %
    %
    %
    %
    %
    %
    %
    %
    %
    %
    \newcount\referencecount \referencecount=0
    \newcount\refsequence \refsequence=0
    \newcount\lastrefno   \lastrefno=-1
    %
    \def\NPrefs{\let\refmark=\NPrefmark \let\refitem=\NPrefitem}
    
    %
    \def\refsymbol#1{\refrange#1-\end}%
    \def\[#1]#2%
          {%
          \if.#2\rlap.\refmark{\refsymbol{#1}}\let\refendtok=\relax%
          \else\if,#2\rlap,\refmark{\refsymbol{#1}}\let\refendtok=\relax%
          \else\refmark{\refsymbol{#1}}\let\refendtok=#2\fi\fi%
          \discretionary{}{}{}\refendtok}%
    \def\refrange #1-#2\end%
        {%
        \refnums #1,\end
        \def \temp {#2}%
        \ifx \temp\empty \else -\expandafter\refrange \temp\end \fi
        }%
    \def\refnums #1,#2\end%
        {%
        \def \temp {#1}%
        \ifx \temp\empty \else \skipspace \temp#1\end\fi
        \ifx \temp\empty
          \ifcase \refsequence
              \or\or ,\number\lastrefno
                \else  -\number\lastrefno
          \fi
          \global\lastrefno = -1
          \global\refsequence = 0
        \else
          \RefObj \edef\temp {Ref_\temp\space}%
          \expandafter \LookUp \temp \using\referencecount\SaveObject
          \global\advance \lastrefno by 1
          \edef \temp {\number\lastrefno}%
          \ifx \label\temp
              \global\advance\refsequence by 1
          \else
              \global\advance\lastrefno by -1
              \ifcase \refsequence
                  \or ,%
                  \or ,\number\lastrefno,%
              \else   -\number\lastrefno,%
              \fi
              \label
              \global\refsequence = 1
              \ifx\suffix\empty
                  \global\lastrefno = \label
              \else
                    \global\lastrefno = -1
              \fi
          \fi
          \refnums #2,\end
        \fi
        }%
    %
    %
    %
    \def\refnum #1{\RefObj \LookUp Ref_#1
                 \using\referencecount\SaveObject \label}%
    \def\reflist  {\RefObj \ListObjects}%
    \def\Refer #1{Ref.[\refsymbol{#1}]}%
    %
    %
    \newif\ifSaveFile
    \newif\ifnotskip
    \newwrite\SaveFile
        \let\IfSelect=\iftrue
    \edef\savefilename {\jobname.aux}%
    \def\Def#1#2%
        {%
        \expandafter\gdef\noexpand#1{#2}%
        \DefObj \SaveObject {#2}{\expandafter\gobble\string#1}%
    }%
    \def\savestate%
        {%
        \ifundefined {chapternumber} \else
          \NumObj \SaveObject {\number\chapternumber}{chapternumber} \fi
            \ifundefined {appendixnumber} \else
          \NumObj \SaveObject {\number\appendixnumber}{appendixnumber} \fi
        \ifundefined {sectionnumber} \else
          \NumObj \SaveObject {\number\sectionnumber}{sectionnumber} \fi
        \ifundefined {pagenumber} \else
          \advance\pagenumber by 1
          \NumObj \SaveObject {\number\pagenumber}{pagenumber}%
          \advance\pagenumber by -1 \fi
        \NumObj \SaveObject {\number\equanumber}{equanumber}%
        \NumObj \SaveObject {\number\tablecount}{tablecount}%
            \NumObj \SaveObject {\number\figurecount}{figurecount}%
        \NumObj \SaveObject {\number\theoremcount}{theoremcount}%
        \NumObj \SaveObject {\number\referencecount}{referencecount}%
        \checkchapterlabel
        \ifundefined {chapterlabel} \else
          {\protect\xdef\chaplabel{\chapterlabel}}
          \MiscObj \SaveObject \chaplabel {chapterlabel} \fi
        \ifundefined {chapterstyle} \else
          \StyleObj \SaveObject
               {\stylename{\chapterstyle}}{chapterstyle} \fi
        \ifundefined {appendixstyle} \else
          \StyleObj \SaveObject
               {\stylename{\appendixstyle}}{appendixstyle}\fi
    }%
    \def\Contents #1{\ObjClass=-#1 \SaveContents}%
    \def\Define #1#2#3%
        {%
        \ifnum #1=\ClassNum
          \global \csname#2\endcsname = #3 %
        \else \ifnum #1=\ClassStyle
          \global \csname#2\endcsname\expandafter=
            \expandafter{\csname#3\endcsname} %
        \else \ifnum #1=\ClassDef
            \expandafter\gdef\csname#2\endcsname{#3} %
        \else
          \expandafter\xdef \csname#2\endcsname {#3} \fi\fi\fi %
        \ObjClass=#1 \SaveObject {#3}{#2}%
        }%
    \def\SaveObject #1#2%
        {%
        \ifSaveFile \else \OpenSaveFile \fi
        \immediate\write\SaveFile
          {%
          \noexpand\IfSelect\noexpand\Define
          {\the\ObjClass}{#2}{#1}\noexpand\fi
          }%
        }%
    \def\SaveContents #1%
        {%
        \ifSaveFile \else \OpenSaveFile \fi
        \BreakLine
        \SaveLine {#1}%
        }%
    \begingroup
            \catcode`\^^M=\active %
    \gdef\BreakLine %
        {%
        \begingroup %
        \catcode`\^^M=\active %
        \newlinechar=`\^^M %
        }%
    \gdef\SaveLine #1#2%
        {%
        \toks255={#2}%
        \immediate\write\SaveFile %
          {%
          \noexpand\IfSelect\noexpand\Contents
          {-\the\ObjClass}{#1}\LBrace\the\toks255\RBrace\noexpand\fi%
          }%
        \endgroup %
        }%
    \endgroup
    \def\ListObjects #1%
        {%
        \ifSaveFile \CloseSaveFile \fi
        \let \IfSelect=\GetContents \ReadFileList #1,\savefilename,\end
           \let \IfSelect=\IfDoObject  \input \savefilename
        \let \IfSelect=\iftrue
        }%
    \def\ReadFileList #1,#2\end%
        {%
        \def \temp {#1}%
        \ifx \temp\empty \else \skipspace \temp#1\end \fi
        \ifx \temp\empty \else \input #1 \fi
        \def \temp {#2}%
        \ifx \temp\empty \else \ReadFileList #2\end \fi
        }%
    \def\GetContents #1#2#3%
        {%
        \notskipfalse
        \ifnum \ObjClass=-#2
          \expandafter\ifx \csname #3\endcsname
            \relax \else \notskiptrue \fi
        \fi
        \ifnotskip \expandafter \DefContents \csname #3_\endcsname
        }%
    \def\DefContents #1#2{\toks255={#2} \xdef #1{\the\toks255}}%
        \def\IfDoObject #1#2%
        {%
        \notskipfalse \ifnum \ObjClass=#2
            \notskiptrue\fi \ifnotskip \DoObject
        }%
    \def\DoObject #1#2%
        {%
        \ifnum \ObjClass = \ClassTbl      \par\noindent Table~#2.
        \else \ifnum \ObjClass = \ClassFig        \par\noindent Figure~#2.
        \else \ifnum \ObjClass = \ClassRef  \refitem{#2}
        \else \item {#2.}
        \fi\fi\fi
        \ifdraft\edef\temp
           {\trimprefix #1\end}[\expandafter\gobble \temp]~\fi
        \expandafter\ifx \csname #1_\endcsname \relax
          \ifdraft\relax\else\edef\temp {\trimprefix #1\end}%
          [\expandafter\gobble \temp]\fi%
        \else
          \csname #1_\endcsname
        \fi
        }%
    \def\OpenSaveFile   {\immediate\openout\SaveFile=\savefilename
                       \global\SaveFiletrue}%
    \def\CloseSaveFile  {\immediate\closeout
                     \SaveFile \global\SaveFilefalse}%
    %
    %
    \def\LookUp #1 #2\using#3#4%
        {%
        \expandafter \ifx\csname#1\endcsname \relax
            \global\advance #3 by 1
            \expandafter \xdef \csname#1\endcsname {\number #3}%
            \let \newlabelfcn=#4%
              \ifx \newlabelfcn\relax \else
               \expandafter \newlabelfcn \csname#1\endcsname {#1}%
             \fi
        \fi
        \xdef \label  {\csname#1\endcsname}%
        \gdef \suffix {#2}%
        \ifx \suffix\empty \else
            \xdef \suffix {\expandafter\trimspace \suffix\end}%
            \xdef \label  {\label\suffix}%
        \fi
        }%
       %
    %
    %
    \newcount\appendixnumber        \appendixnumber=0
    \newtoks\appendixstyle                \appendixstyle={\Alphabetic}
    \newif\ifappendixlabel                \appendixlabelfalse
    \def\APPEND#1{\par\penalty-300\vskip\chapterskip
          \spacecheck\chapterminspace
          \global\chapternumber=\number\appendixnumber
          \global\advance\appendixnumber by 1
          \chapterstyle\expandafter=\expandafter{\the\appendixstyle}
    \chapterreset
         \titlestyle{Appendix\ifappendixlabel~\chapterlabel\fi.~ {#1}}
          \nobreak\vskip\headskip\penalty 30000}
    %

    %
    %
    %
    \def\references#1{\par\penalty-300\vskip\chapterskip\spacecheck
           \chapterminspace\line{\fourteenrm\hfil References\hfil}
           \nobreak\vskip\headskip\penalty 30000\reflist{#1}}
    \def\figures#1{\par\penalty-300\vskip\chapterskip\spacecheck
         \chapterminspace\line{\fourteenrm\hfil Figure Captions\hfil}
         \nobreak\vskip\headskip\penalty 30000\figlist{#1}}
    \def\tables#1{\par\penalty-300\vskip\chapterskip\spacecheck
          \chapterminspace\line{\fourteenrm\hfil Table Captions\hfil}
           \nobreak\vskip\headskip\penalty 30000\tbllist{#1}}
    \newif\ifdraft\draftfalse
    \newcount\yearltd\yearltd=\year\advance\yearltd by -1900
    \def\draft{\drafttrue
        \def\draftdate{preliminary draft:
            \number\month/\number\day/\number\yearltd\ \ \hourmin}%
            \paperheadline={\hfil\draftdate} \headline=\paperheadline
            {\count255=\time\divide\count255 by 60
                       \xdef\hourmin{\number\count255}
                \multiply\count255 by-60\advance\count255 by\time
        \xdef\hourmin{\hourmin:\ifnum\count255<10 0\fi\the\count255} }
          \message{draft mode}  }
    %
    \def\slacpub{ \Pubnum
={$\caps SLAC - PUB - \the\pubnum $}}
    %
    %
    
    \def\frac#1/#2{\leavevmode\kern.1em\raise.5ex
                  \hbox{\the\scriptfont0
                  #1}\kern-.1em/\kern-.15em
                  \lower.25ex\hbox{\the\scriptfont0 #2}}
    %
    %

    \def\frac#1#2{{#1 \over #2}}
    
    \def\half{{\frac 12}}

    \def\eq#1{$\(#1)$}
    %

\IfSelect \Contents
{-3}{Ref_thornmosc}{C. B. Thorn,
``Reformulating String Theory with the 1/N Expansion,''
in {\it Sakharov Memorial Lectures in Physics},
Ed. L. V. Keldysh and V. Ya. Fainberg, Nova Science Publishers Inc.,
Commack, New York, 1992. hep-th/9405069}\fi
\IfSelect \Contents
{-3}{Ref_susskindtu}{L. Susskind, L. Thorlacius, and J. Uglum,
{\sl Phys. Rev.} {\bf D48} (1993) 3743.}\fi
\IfSelect \Contents
{-3}{Ref_klebanovs}{I. Klebanov and L. Susskind,
{\sl Nucl. Phys.} {\bf B309} (1988) 175.}\fi
\IfSelect \Contents
{-3}{Ref_susskindbh}{L. Susskind,
``Some Speculations About Black Hole Entropy in String Theory,''
Rutgers Univ. preprint RU-93-44, hep-th/9309145;
L. Susskind and J. Uglum, ``Black Hole Entropy in Canonical
Quantum Gravity and Superstring Theory,'' Stanford Univ. preprint,
}\fi
\IfSelect \Contents
{-3}{Ref_thooftlargen}{G. 't Hooft,
{\sl Nucl. Phys.} {\bf B72} (1974) 461.}\fi
\IfSelect \Contents
{-3}{Ref_thooftbh}{G. 't Hooft,
{\sl Nucl. Phys.} {\bf B342} (1990) 471;
``On the Quantization of Space and Time,'' {\it Proc. of the
4th Seminar on Quantum Gravity}, 25\dash29 May 1987, Moscow, USSR,
ed. M. A. Markov \etal, (World Scientific Press, 1988);
``Dimensional Reduction in Quantum Gravity,'' Utrecht preprint,
THU-93/26, GR-QC/9310026.}\fi
\IfSelect \Contents
{-3}{Ref_weinbergw}{S. Weinberg and E. Witten,
{\sl Phys. Lett.} {\bf 96B} (1980) 59.}\fi
\IfSelect \Contents
{-3}{Ref_nielsenfishnet}{H. B. Nielsen and P. Olesen,
{\sl Phys. Lett.} {\bf 32B} (1970) 203;
B. Sakita and M. A. Virasoro,
{\sl Phys. Rev. Lett.} {\bf 24} (1970) 1146.}\fi
\IfSelect \Contents
{-3}{Ref_gilest}{R. Giles and C. B. Thorn,
{\sl Phys. Rev.} {\bf D16} (1977) 366.}\fi
\IfSelect \Contents
{-3}{Ref_thornfishnet}{C. B. Thorn,
{\sl Phys. Rev.} {\bf D17} (1978) 1073.}\fi
\IfSelect \Contents
{-3}{Ref_bardakcis}{K. Bardakci and S. Samuel,
{\sl Phys. Rev.} {\bf D16} (1977) 2500.}\fi
\IfSelect \Contents
{-3}{Ref_gilesmt}{R. Giles, L. McLerran, C. B. Thorn,
{\sl Phys. Rev.} {\bf D17} (1978) 2058.}\fi
\IfSelect \Contents
{-3}{Ref_thornfock}{C. B. Thorn,
{\sl Phys. Rev.} {\bf D20} (1979) 1435.}\fi
\IfSelect \Contents
{-3}{Ref_goddardgrt}{P. Goddard, J. Goldstone,
C. Rebbi, and C. B. Thorn, {\sl Nucl. Phys.} {\bf B56} (1973) 109.}\fi
\IfSelect \Contents
{-3}{Ref_kakukikkawa}{M. Kaku and K. Kikkawa,
{\sl Phys. Rev.} {\bf D10} (1974) 1823.}\fi
\IfSelect \Contents
{-3}{Ref_thornsantab}{C. B. Thorn, in
{\it Unified String Theories,} ed. M. Green and D. Gross,
World Scientific Publishing Co. (1986).}\fi
\IfSelect \Contents
{-3}{Ref_thornweeparton}{C. B. Thorn,
{\sl Phys. Rev.} {\bf D19} (1979) 639.
See also I. Klebanov and L. Susskind,
{\sl Nucl. Phys.} {\bf B309} (1988) 175, for a
lattice model with similar physics.}\fi
\IfSelect \Contents
{-3}{Ref_zeldovich}{Ya. B. Zel'dovich,
{\sl Zh. Eksp. Teor. Fiz. Pis'ma Red} {\bf 6} (1967) 883
[{\sl JETP Lett} {\bf 6} (1967) 316].}\fi
\IfSelect \Contents
{-3}{Ref_sakharov}{A. D. Sakharov,
{\sl Dok. Akad. Nauk. SSSR} {\bf 177} (1967) 70
[{\sl Sov. Phys. Dokl.} {\bf 12} (1968) 1040].}\fi
\IfSelect \Contents
{-3}{Ref_adlerinducedgrav}{S. L. Adler,
{\sl Phys. Rev.} {\bf D14} (1976) 379; {\sl Phys. Rev. Lett.}
{\bf 44} (1980) 1567; {\sl Phys. Lett.}
{\bf 95B} (1980) 241; {\sl Rev. Mod. Phys.}
{\bf 54} (1982) 729.}\fi
\IfSelect \Contents
{-3}{Ref_zeeinducedgrav}{A. Zee,
{\sl Phys. Rev. Lett.} {\bf 42} (1979) 417;
{\sl Phys. Rev.} {\bf D23} (1981) 858;
{\sl Phys. Rev. Lett.} {\bf 48} (1982) 295;
{\sl Phys. Lett.} {\bf 109B} (1982) 183.}\fi
\IfSelect \Contents
{-3}{Ref_wittentop}{E. Witten,
{\sl Comm. Math. Phys.} {\bf 117} (1988) 353.}\fi
\IfSelect \Contents
{-3}{Ref_woodardunstablesft}{D. A. Eliezer and R. P. Woodard,
{\sl Nucl. Phys.} {\bf B325} (1989) 389.}\fi
\IfSelect \Contents
{-3}{Ref_kazakov}{V. A. Kazakov,
{\sl Mod. Phys. Lett.} {\bf A4} (1989) 2125.}\fi
\IfSelect \Contents
{-3}{Ref_kazakovm}{V. A. Kazakov and A. A. Migdal,
{\sl Nucl. Phys.} {\bf B311} (1988/89) 171.}\fi
\IfSelect \Contents
{-3}{Ref_brezink}{E. Brezin and V. A. Kazakov,
{\sl Phys. Lett.} {\bf 236B} (1990) 144.}\fi
\IfSelect \Contents
{-3}{Ref_grossm}{D. J. Gross and A. A. Migdal,
{\sl Phys. Rev. Lett.} {\bf 64} (1990) 127;
{\sl Phys. Rev. Lett.} {\bf 64} (1990) 717;
{\sl Nucl. Phys.} {\bf B340} (1990) 333.
{\sl Phys. Rev. Lett.} {\bf 64} (1990) 717.
}\fi
\IfSelect \Contents
{-3}{Ref_douglass}{M. R. Douglas and S. H. Shenker,
{\sl Nucl. Phys.} {\bf B335} (1990) 635.}\fi
\IfSelect \Contents
{-3}{Ref_grossmende}{D. J. Gross and P. F. Mende,
{\sl Phys. Lett.} {\bf 197B} (1987) 129;
{\sl Nucl. Phys.} {\bf B303} (1988) 407.}\fi
\IfSelect \Contents
{-3}{Ref_atickw}{J. J. Atick and E. Witten,
{\sl Nucl. Phys.} {\bf B310} (1988) 291.}\fi
\IfSelect \Contents
{-3}{Ref_goldstonemany}{J. Goldstone,
{\sl Proc. Roy. Soc. (London)} {\bf A239} (1957) 267.}\fi
\IfSelect \Contents
{-3}{Ref_goldstonepriv}{J. Goldstone,
private communication (1978)}\fi
\pubnum={$94-8$\cr
hep-th/9407169\cr}
\date{}
\pubtype={}
\titlepage
\title{Calculating the Rest Tension for a Polymer
of String Bits\footnote{*}{
Work supported in part by the Department of Energy,
contract DE-FG05-86ER-40272}}
\author{Charles B. Thorn
}
\address{Department of Physics, University of Florida, Gainesville,
FL, 32611, USA }
\abstract
We explore the application of approximation schemes from
many body phys\-ics, including the Hartree-Fock method
and random phase approximation (RPA), to the problem
of analyzing the low energy excitations of a polymer
chain made up of bosonic string bits. We accordingly
obtain an expression for the rest tension $T_0$ of
the bosonic relativistic string in terms of the
parameters characterizing the microscopic string bit
dynamics. We first derive an exact connection between the
string tension and a certain correlation function of
the many-body string bit system. This connection
is made for an arbitrary interaction potential between
string bits and relies on an exact dipole sum rule.
We then review an
earlier calculation by Goldstone of the low
energy excitations of a polymer chain
using RPA. We assess the accuracy of the RPA by
calculating the first order corrections. For this
purpose we specialize to the unique scale invariant
potential, namely an attractive delta function potential
in two (transverse) dimensions. We find that the
corrections are large, and discuss a method for summing
the large terms. The corrections to this improved RPA
are roughly 15\%.
\endpage
\pagenumbers
\chapter{Introduction}
A decade and a half ago we proposed a microscopic theory of
the relativistic string.\[gilest,thornfishnet,thornweeparton]\
At that time the
main motivation was to seek a connection between string
theory and QCD. From today's vantage point this motivation
appears wrong-headed, but the model we proposed then still
holds promise for a fundamental reformulation of
string theory. Some of the same ideas have reemerged in
recent years in the context of large $N$ matrix models,
which have been the subject of vigorous
investigation.\[kazakov,kazakovm,brezink,grossm,douglass]\
In 1991\[thornmosc]\ we sketched why the wee parton model
of Ref.[\refnum{thornweeparton}] is an attractive starting point
for a fundamental formulation of string theory.

In this article
we begin a serious study of the detailed relationship
of the microscopic physics of such a string bit model to
the low energy properties of string. Presumably, the microscopic
details of the model would only be apparent at the Planck
scale. For the purposes of our present analysis we imagine
that the string coupling is small, \ie\ the string mass scale
$\sqrt{T_0}<<M_{Planck}$. Even though this is probably
an unrealistic assumption, it is useful because it is
 a limit which puts highly nontrivial constraints on the
string bit model. In particular, it is a limit in which
the graviton, a composite of string bits in our model,
must satisfy all of the low energy theorems enjoyed by
Einstein's general relativity. These constraints are
guaranteed, of course, once it is demonstrated that
the low energy properties of our string bit system are
precisely those of the relativistic string.

The work of \Refer{thornweeparton} established the
microscopic/macroscopic connection between
string bits and the bosonic string in the context of the Random
Phase Approximation (RPA). It can be argued that
the corrections to RPA can be lumped into a renormalization
of the string tension, and so are not important as far as
the stringy properties of the low energy
theory are concerned. However the
actual size of the corrections to RPA was not
determined in \Refer{thornweeparton}. Since the actual
microscopic details of the string bits may be of eventual
interest (\eg\ in resolving the short distance properties of
gravity), we believe that it is important to assess the
validity of RPA quantitatively, and this is our goal
here. Moreover, the details of the micro/macro connection
are probably essential for a proper string bit understanding
of the superstring. For example, must we require
an exact supersymmetry at the level of string bits, or
is supersymmetry only present in the low energy
theory? We don't answer that question here, but it is
clearly an important one, and the answer is surely
going to require a more extensive treatment than
that given in \Refer{thornweeparton}.

In the following sections we provide this more detailed
study of bosonic string bits. In Section 2, we present the
dynamics of a polymer of string bits in Goldstone's
formulation of the many body problem.\[goldstonemany]\
This formulation provides an elegant presentation of various
``improved'' versions of perturbation theory, such as
Hartree-Fock methods and RPA, in terms of Feynman
diagrams. In Section 3, we use this diagrammatic method
to define certain ``irreducible'' correlation functions,
for which exact low frequency theorems can be derived.
These theorems are then used to show the universal
presence of harmonic low energy excitations, which
reproduce the usual spectrum of the relativistic string.
Thus the string tension will be related to the
value a certain
zero-frequency correlation function
$\int dt\bra{G}Tx^r(t)x^s(0)\ket{G}$.
This holds even for string bit interaction potentials
that are not differentiable or even finite at minimum
separation. A case in point is the potential
$V({\bf x})=-\lambda_0\delta({\bf x})$, which has the
nice property of being scale invariant in two
dimensions\dash precisely the case for the transverse
space of four-dimensional space-time.
But evaluation of the relevant correlation function
requires an exact solution of the string bit system,
which we don't have except in the case of a
harmonic oscillator string bit potential. Our
string bit model involves a short range, scale invariant potential.
This is where we must resort to approximations. In
Section 4, we review Goldstone's application of RPA
to the string bit polymer problem and then proceed
to calculate the first order corrections to RPA.
We find that the diagrams which have the character
of ``self-energy'' corrections are in fact quite
large. Fortunately the structure of these large
corrections is simple enough that they can be summed
to all orders, which we do in Section 5. The upshot is
that the large corrections have the effect of
renormalizing the ionization energy of the polymer,
and once this is taken into account the remaining
corrections are small, of order 15\%. Finally
we gather some conclusions in Section 6.

\chapter{Chain Dynamics in Formalism of Second Quantization}
The model of string bits proposed in
\Refer{thornweeparton,thornmosc}
leads to the relativistic string in light-cone gauge\[goddardgrt].
Although
the string formed from string bits moves in $D$ space-time
dimensions (described by light-cone coordinates
$$\eqalign{
x^\pm=&(x^0\pm x^1)/\sqrt{2}\cr
{\bf x}=&(x^2,\cdots,x^{d})\cr}
$$
where $d=D-1$ is the spatial dimension), the string
bits move only in the transverse space, with
coordinates ${\bf x}$: in fact they enjoy
a {\it Newtonian} (\ie\ Galilean invariant)
dynamics in this transverse space.
The $x^+\equiv\tau$ coordinate is a Newtonian time for the
string bits, but the longitudinal dimension,
corresponding to $x^-$ and its conjugate
momentum $P^+$, is nonexistent
for them. It only emerges with string
formation: bit number is a conserved quantity
in string bit dynamics, and for a string with a
large number of bits it becomes an effectively
continuous variable, which can be consistently
interpreted as a variable $P^+$. In this interpretation, we
can think of each string bit as carrying a tiny fixed amount
of $P^+=\epsilon$. Then a string containing $M$ string bits
would carry $P^+=M\epsilon$. The fact that the $P^-$ of
the string should vary as $1/P^+$, is then understood as
the $1/M$ dependence of low energy excitations of large systems.
The hamiltonian describing
string bit dynamics will accordingly be identified
as the operator $\epsilon P^-$
for strings with an infinite number of string bits.

The string bit dynamics proposed in \Refer{thornweeparton,thornmosc}
exploits 't Hooft's ideas
on the $1/N$ expansion\[thooftlargen]. We postulated
that the string bit annihilation operator is
an $N\times N$ matrix variable
$a_k^{~\ell}({\bf x})$ and denoted the string bit
creation operator by
${\bar a}_k^{~\ell}({\bf x})=a_\ell^{~k}({\bf x})^\dagger$.
Then
$$[a_k^{~\ell}({\bf x}),{\bar a}_m^{~n}({\bf y})]
=\delta_k^n\delta_m^\ell\delta({\bf x}-{\bf y}).
$$
In the large $N$ limit, one can show that the
singlet operators
$${\bar A}({\bf x}_1,\cdots,{\bf x}_M)=
\left({
{1\over N}}\right)^{M/2}
\tr\{{\bar a}({\bf x}_1)\cdots{\bar a}({\bf x}_M)\}
$$
behave as creation operators
for closed polymers of string bits.\[thornfock]\
Note here that Bose statistics for the string
bits implies cyclic symmetry for the
$\bar A$'s. This symmetry accounts for the
well-known $L_0={\bar L}_0$ constraint
in the eventual low energy string dynamics.

 A candidate dynamics for bosonic string bits is
given by the Hamiltonian
$$\eqalign{
H\equiv&\epsilon P^-=\int d{\bf x}{1\over 2}
\tr{\bf\nabla}{\bar a}({\bf x})\cdot{\bf\nabla}a({\bf x})\cr
&+{1\over2 N}\int
d{\bf x} d{\bf y}{\cal V}({\bf x}-{\bf y})
\tr:({\bar a}({\bf x}){a}({\bf x})-{a}({\bf x}){\bar a}({\bf x}))
({\bar a}({\bf y}){a}({\bf y})-{a}({\bf y}){\bar a}({\bf y})):\cr}
\(constitham)
$$
where ${\cal V}$ should be positive (repulsive) for
stability.
Evaluating the action of $H$ on a state
$$\ket{\psi}=\left({
{1\over N}}\right)^{M/2}
\tr\{{\bar a}({\bf x}_1)\cdots{\bar a}({\bf x}_M)\}
\psi_M({\bf x}_1,\cdots,{\bf x}_M)\ket{0},
$$
leads to
$$\eqalign{
H\ket{\psi}&=
{\textstyle|\sum_i^M(-{\bf\nabla_i}^2/2
-{\cal V}({\bf x}_{i+1}-{\bf x}_i))\psi\rangle}\cr
-&{1\over N}\sum_{i=1}^M\sum_{j\neq i,i+1}
{\bar A}({\bf x}_{i+1},\cdots,{\bf x}_{j-1})
{\bar A}({\bf x}_{j},\cdots,{\bf x}_{i})\ket{0}
{\cal V}({\bf x}_i-{\bf x}_j)
\psi_M({\bf x}_1,\cdots,{\bf x}_M)\cr
&+{\rm Other~Terms~of~Order~(1/N)}.\cr}
\(haction)
$$
The displayed $1/N$ term describes the dissociation of a
closed polymer into two closed polymers,
so $1/N$ can be identified with the
string coupling constant. In the limit that it
vanishes ($N\rightarrow\infty$),
the above equation describes a stable
polymer provided that the nearest neighbor potential $-{\cal V}$
is sufficiently attractive to bind a pair of
string bits\[thornweeparton]. We shall review the argument
below.

Our aim in this article is to go as far as we can
toward the solution of this large $N$ limit of the string
bit system. Referring to \eq{haction} we see that the
energy eigenstates in this limit are determined by the
solution of the following nonrelativistic many body
Schr\"odinger equation:
$$\sum_{i=1}^M(-{\bf\nabla_i}^2/2
-{\cal V}({\bf x}_{i+1}-{\bf x}_i))\psi({\bf x_1},\ldots,{\bf x_M})
=E\psi({\bf x_1},\ldots,{\bf x_M}).\(manybodyprob)$$
For purposes of keeping this article reasonably
self-contained, the rest of this section
will be devoted to a recapitulation
of the main results already reported in \Refer{thornweeparton}.
New developments will be reserved for the subsequent sections.

It is first convenient to use Galilean invariance to separate
the motion of the center of mass. So define
$$\eqalign{
{\bf X}\equiv& {1\over M}\sum_r{\bf x}_r\cr
{\bf y}_r\equiv& {\bf x}_r-{\bf x}_{r-1}\cr}
\(cmseparation)
$$
where ${\bf x}_0\equiv {\bf x}_M$ and ${\bf x}_{M+1}\equiv{\bf x}_1$.
Note that although we have defined a ${\bf y}_r$ for each
$r$, only $M-1$ of them are independent: they satisfy the
constraint
$${\bf Y}\equiv{1\over M}\sum_{r=1}^M {\bf y}_r=0.$$
Choosing ${\bf y}_2,\ldots, {\bf y}_M$ as independent
variables, and denoting $(1/i)\partial/\partial{\bf X}\equiv{\bf P}$
and $(1/i)\partial/\partial{\bf y}_r\equiv{\bf p}_r$, the
many body Hamiltonian corresponding to \eq{manybodyprob}
becomes
$$\eqalign{
h=&{{\bf P}^2\over2M}
+\sum_{r=2}^M {\bf p}_r^2-\sum_{r=2}^{M-1}{\bf p}_r\cdot{\bf p}_{r+1}
-\sum_{r=2}^M{\cal V}({\bf y}_r)-{\cal V}(-{\bf y}_2-\cdots-{\bf y}_M)\cr
\equiv&{{\bf P}^2\over2M}+h^\prime.\cr}\(polymerham)
$$
The first term on the r.h.s., giving the energy of center of mass
motion, is precisely what is needed for the consistent interpretation
of $M\epsilon$ as $P^+$ and $h/\epsilon$ as $P^-$. The remainder
of the r.h.s. $h^\prime$ should have the interpretation (for $M\rightarrow
\infty$) as ${\cal M}^2/2M$ where ${\cal M}$ is the rest mass operator
for the string. Since there will inevitably be a bulk contribution
to the eigenvalues of $h^\prime$ proportional to the number of
string bits $M$, we must subtract this contribution before
the $1/M$ behavior is revealed. This subtraction is easily
accomplished by including a counterterm in the original
string bit Hamiltonian of the form $\mu_0\int d{\bf x}\tr\bar a({\bf x})
a({\bf x})$. Note that in the low energy theory such a term is
just a multiple of $P^+$. The remainder of this paper is devoted
entirely to gaining an understanding of the eigenspectrum of $h^\prime$
especially the excitation energies of order $(1/M)$ above the ground
state energy.

We first use variational arguments to establish a simple criterion
for the stability of the string bit polymers. Here we are assessing
stability against the breaking of nearest neighbor bonds, \ie\
the dissociation into a number of open polymers with all
unbroken bonds retaining their arrangement. Dissociation of
a closed polymer into two smaller closed polymers by bond
rearrangement is another matter intimately associated with
the presence or absence of tachyons in the low energy string
theory. Such decays are vetoed by the $N\rightarrow\infty$
limit corresponding to zero string coupling constant.

We assume that the potential $-{\cal V}$ is negative and vanishes at
large distances. The zero of energy corresponds to all the string
bits infinitely dispersed from one another. Notice that
deleting the last term in \eq{polymerham} leaves the Hamiltonian
for an open polymer with the same number of bits. Since the
deleted term is negative, this shows that the closed polymer
ground state energy is strictly less than the open polymer
ground state energy. Thus stability of the open polymer implies
that of the closed polymer. Next consider the dissociation
of an open polymer with $M$ bits into two open polymers with
$M_1$ and $M-M_1$ bits respectively. Because of the
nearest neighbor interaction structure, the internal Hamiltonian
for an open polymer can be broken down as:
$$h^\prime_{{\rm Open},M}=h^\prime_{{\rm Open}, M_1}
+h^\prime_{{\rm Open}, M-M_1}
+{\bf p}_{M_1+1}^2-{\cal V}({\bf y}_{M_1+1})
-{\bf p}_{M_1}\cdot{\bf p}_{M_1+1}-{\bf p}_{M_1+1}\cdot
{\bf p}_{M_1+2},
$$
where the variables labelled by $M_1+1$ do not appear in either
of the open polymer Hamiltonians on the r.h.s. Thus applying
the variational principle with a trial wave function given
by $\psi_{Trial}=\psi^G_{M_1}\psi^G_{M-M_1}\psi^G_2({\bf y}_{M_1+1})$,
where $\psi^G_K$ is the exact ground state wave function for
a polymer with $K$ bits, gives
$$E_{{\rm Open}, M}^G\leq E_{{\rm Open}, M_1}^G
+E_{{\rm Open}, M-M_1}^G+E^G_2.$$
Here one uses the fact that $\VEV{{\bf p}_r}=0$ in any of
the ground state wave functions in the trial.
If the nearest neighbor potential binds two bits, $E^G_2<0$
and the ionization energy required for
dissociation into two open polymers
is {\it at least} $B=-E^G_2$.

Now for large $M$, the ground state energy of an open
polymer should have the behavior
$$E^G_{{\rm Open}, M}\sim
\alpha M + \beta + \gamma/M +\cdots$$
Inserting this form into the variational estimate gives
the rigorous bound $\beta\geq-E^G_2=B$. More precisely
we see that $\beta$ has the interpretation as the ionization
energy required to break a long open polymer into two smaller
but still long open polymers. This ionization energy
is just the energy cost of breaking a bond. Since the
transformation of a closed polymer to an open polymer
requires breaking precisely one bond, and for very long
polymers the energy cost of this break can only differ
 from that of breaking an open polymer into two open
polymers by energies of order $1/M$, we can conclude that
the ground state energy of a closed polymer should have
the behavior
$$E^G_{{\rm Closed},M}\sim \alpha M + \gamma^\prime/M.$$
The coefficients of the $1/M$ terms are, of course,
different in the open and closed cases. It is crucial
for the interpretation of $h$ as $\epsilon P^-$ that
there is no constant $\beta$ term for the closed polymer
ground state energy, since only the
bulk $\alpha$ term can be removed
by a counterterm in the string bit Hamiltonian.

The important parameter that characterizes the stringy
properties of long polymers is the string rest tension
$T_0$. It can be inferred from the low lying level spacing
of the polymers. Eigenvalues of $h^\prime$ of order $1/M$
above the lowest eigenvalue will correspond to values of
$P^-$ behaving as $1/P^+$. The coefficient of $1/M$ is
thus $2\times{\cal M}^2$. To calculate $T_0$ it is
simplest and sufficient to study long open polymers,
and we shall make this restriction in what follows.

We only remark that the best way to apply the methods
to the closed case is to consider a related system
to \eq{polymerham}, in which one new degree of freedom
$({\bf y}_1, {\bf p}_1)$ is introduced, and the
Hamiltonian is taken to be
$${\hat h}={{\bf P}^2\over2M}
+\sum_{r=1}^M {\bf p}_r^2-\sum_{r=1}^{M}{\bf p}_r\cdot{\bf p}_{r+1}
-\sum_{r=1}^M{\cal V}({\bf y}_r).
$$
To apply solutions of this system to \eq{polymerham}, notice
that the quantity ${\bf Y}$, defined below
\eq{cmseparation}, commutes with
${\hat h}$. One can therefore look for eigenfunctions of
${\hat h}$ of the form $\delta({\bf Y})
\psi({\bf y}_2,\ldots,{\bf y}_M)$,
whence $\psi$ must be an eigenfunction of \eq{polymerham}.
Thus the eigenstates of \eq{polymerham} are the subset of
eigenstates of ${\hat h}$ which satisfy the constraint ${\bf Y}=0$.

Henceforth we restrict attention to the open polymer hamiltonian
with center of mass motion removed
$$h^\prime_{{\rm Open}}\equiv
\sum_{r=2}^M {\bf p}_r^2-\sum_{r=2}^{M-1}{\bf p}_r\cdot{\bf p}_{r+1}
-\sum_{r=2}^M{\cal V}({\bf y}_r).\(openpolymerham)
$$
Our variational arguments have already established that the
ground state of this system is a discrete bound state lying at
least an amount $B$ below the continuum. We must next look
for discrete levels lying $O(1/M)$ above the ground level.
For this purpose, we apply many body methods developed
by Goldstone\[goldstonemany,goldstonepriv].

Goldstone employs a second quantization formalism
in which a fermionic field $\psi_r({\bf y})$ is introduced
for each bond $r=2,\ldots,M$ in the open polymer chain.
Each bond number
operator $M_r=\int d{\bf y}\psi^\dagger_r({\bf y})\psi_r({\bf y})$
will be conserved by the second quantized dynamics and so
can be consistently set to unity. In this sector the
second quantized Hamiltonian
$$\eqalign{
H_{{\rm Open}}=&\int d{\bf y}\sum_{r=2}^M\psi^\dagger_r
(-\nabla^2-{\cal V}({\bf y}))\psi_r\cr
\quad&-\sum_{r=2}^{M-1}\int d{\bf y}\psi^\dagger_r({\bf y})
(-i\nabla)\psi_r({\bf y})\int d{\bf z}\cdot\psi^\dagger_{r+1}({\bf z})
(-i\nabla)\psi_{r+1}({\bf z})\cr}\(secondq)
$$
is completely equivalent to \eq{openpolymerham}. This
Hamiltonian can now be analyzed in the
interaction picture using Feynman diagrams in
which the propagator is determined by the quadratic term
and the vertex is determined by the quartic term.
The conceptual advantage of this formalism is that
the zeroth order ground state, namely the state in
which each bond is in the ground state of its one
body potential, is simply
expressed as a boundary condition on the propagator,
just as in the Dirac hole theory of positrons.

Expanding
$\psi_r$ in eigenfunctions of the one body hamiltonian
${\bf p}^2-{\cal V}({\bf y})$
$$\psi_r({\bf y})=\sum_{m>0}b^r_m\phi_m({\bf y})+b_0^r\phi_0({\bf y}),$$
where $m=0$ labels the ground state and $m>0$ labels
the excited states, the zeroth order ground state is
specified by
$$b^r_m\ket{0}=0\quad{\rm for}~m>0\qquad b^{r\dagger}_0\ket{0}=0.$$
Then the interaction picture propagator, represented by
the first diagram in \Fig{rules}, is given by
$$\eqalign{
\bra{0}T[\psi_r^I({\bf y},t)&\psi_s^{I\dagger}({\bf z},0)]\ket{0}\cr
=&\delta_{rs}\{\theta(t)\sum_{m>0}\phi_m({\bf y})\phi^*_m({\bf z})
e^{-itE_m}-\theta(-t)\phi_0({\bf y})\phi_0^*({\bf z})e^{-itE_0}\}\cr
=&\delta_{rs}\int{d\omega\over 2\pi}e^{-i\omega t}
\left\{\sum_{m>0}{-i\phi_m({\bf y})\phi^*_m({\bf z})\over
E_m-\omega-i\epsilon}
+{-i\phi_0({\bf y})\phi_0^*({\bf z})\over E_0-\omega+i\epsilon}\right\}.\cr
}
\(propagator)
$$
\insertfigure{rules}{\centerline{\psfig{figure=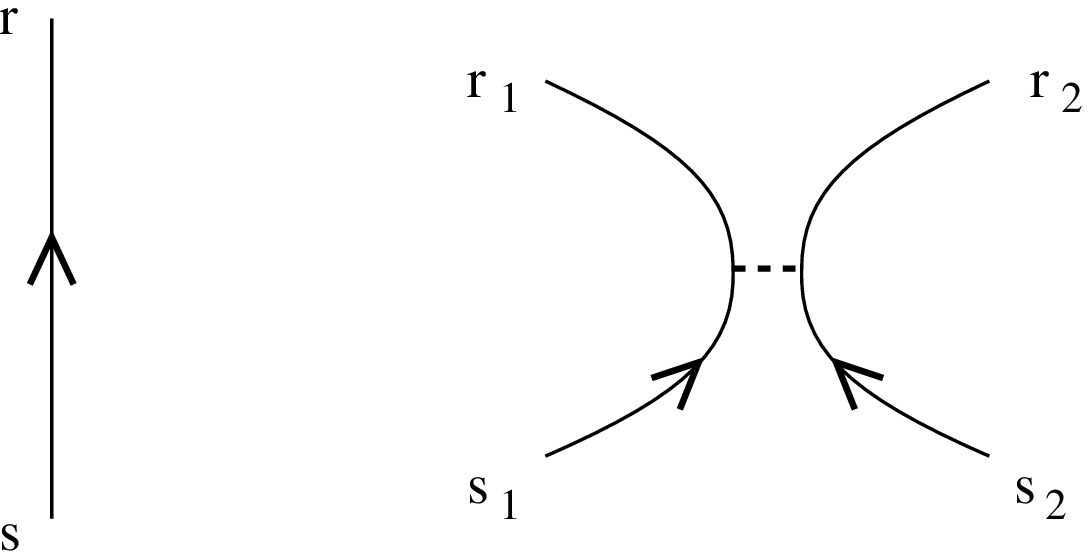,height=4cm}}}
{Feynman rules for a string bit chain.}{\vbox{\vskip4cm}}

The vertex function represented by the second diagram in
\Fig{rules} is given by
$$i\delta_{r_1s_1}\delta_{r_2s_2}(\delta_{r_1,r_2+1}+\delta_{r_1,r_2-1})
(-i\nabla_1)(-i\nabla_2)\(vertex)$$
where the gradient operators act on the first arguments of the
respective propagators directed into the vertex, and
a Kronecker delta with either index equal to 1 or $M+1$ is
understood to vanish.

In order to apply time dependent perturbation theory to
the calculation of ground state expectation values of
time ordered products of operators like ${\bf y}_r$
or ${\bf p}_r$, we simply use Dyson's formula after
translating to the second quantized formalism. Thus
we have
$$\eqalign{
{\bf y}_r\equiv&\int d{\bf y}\psi^\dagger_r({\bf y})
{\bf y}\psi_r({\bf y})\cr
{\bf p}_r\equiv&\int d{\bf y}\psi^\dagger_r({\bf y})
{1\over i}{\partial\over\partial{\bf y}}\psi_r({\bf y})\cr}
$$
so, for example,
$$\eqalign{
&\bra{G}T[y^k_r(t)y^m_s(0)]\ket{G}=\crr
&\qquad\qquad{\bra{0}T[\int d{\bf y}\psi^{I\dagger}_r({\bf y},t)
{ y^k}\psi_r^I({\bf y},t)\int d{\bf y}\psi^{I\dagger}_s({\bf z},0)
{ z^m}\psi_r^I({\bf z},0)
e^{-i\int_{-\infty}^{+\infty}dt^\prime H^{\prime}_I(t^\prime)}]
\ket{0}\over\bra{0}
T[e^{-i\int_{-\infty}^{+\infty}dt^\prime H^{\prime}_I(t^\prime)}]
\ket{0}}\cr}
\(dysonformula)
$$
where $H^\prime_I(t^\prime)$ is just the quartic term of
\eq{secondq} transformed to interaction picture. As usual
the denominator in \eq{dysonformula} simply cancels all
disconnected bubble diagrams and correctly normalizes
the matrix element after the infinite time integral projects
onto the true ground state of the many body system. Information
on the spectrum of the exact Hamiltonian can be inferred
{} from the time dependence of such correlation functions,
or equivalently from the pole locations in their time
Fourier transforms.

By applying the Random Phase Approximation corresponding
to summing the bubble diagrams shown in \Fig{rpaapprox},
Goldstone demonstrated the existence of order $1/M$
excitations in this approximation
and gave an approximate formula for the
string tension. In the following section we put these
insights into an all orders context, and so obtain
an exact formula for the string tension.

\insertfigure{rpaapprox}
{\centerline{\psfig{figure=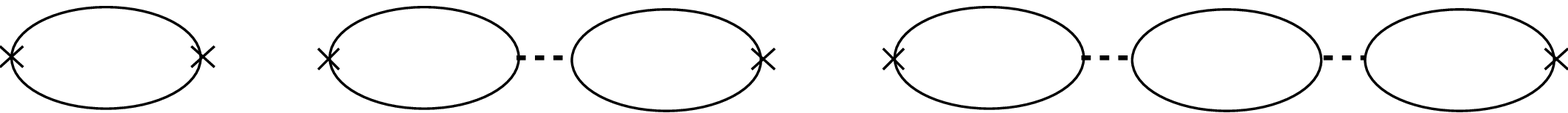,height=1cm}}}
{Diagrams summed in the random phase approximation}{\vbox{\vskip1cm}}

\chapter{Low Energy Excitations: The String Tension}

Consider the system
$$h^\prime_{{\rm Open}} = {\sum_{r=2}^{M} [{\bf p}_r^2 + {\cal V}({\bf y}_r)]}-
{\sum_{r=2}^{M-1} {\bf p}_r\cdot {\bf p}_{r+1}} $$
Our aim is to calculate the low lying excitations of such a system under
the assumption that the system binds together, i.e. that there is a discrete
ground state with a mass gap.  We shall examine the poles in the
time Fourier
transforms of various correlation functions:
$$\eqalign{
\bra{G} T[y_r^i (t) y_s^j(0)]\ket{G} \equiv& - \int {d\omega\over
2\pi i} {\cal E}_{rs}^{ij} (\omega) e^{-i\omega t}\cr
\bra{G} T[p_r^i (t) p_s^j (0)]\ket{G} \equiv& - \int {d\omega\over
2\pi i}{\cal G}_{rs}^{ij} (\omega)e^{-i\omega t}\cr
\bra{G} T [y_r^i (t) p_s^j (0)]\ket{G} \equiv& - \int {d\omega\over
2\pi i}{\cal F}_{rs}^{yp,ij}(\omega)e^{-i\omega t}\cr
\bra{G} T [p_r^i (t) y_s^j (0)]\ket{G} \equiv& - \int {d\omega\over
2\pi i}{\cal F}_{rs}^{py,ij}
(\omega)e^{-i\omega t}.\cr}$$
Using time translation invariance, we can transfer the
time argument to the other member of the time ordered product,
which leads to the identities
$$\eqalign{{\cal G}_{rs} (-\omega) &= {\cal G}_{sr} (\omega)\cr
{\cal E}_{rs} (-\omega) &= {\cal E}_{sr} (\omega)\cr
{\cal F}_{rs}^{py} (-\omega) &= {\cal F}_{sr}^{yp} (\omega).\cr}$$
Using the Heisenberg equations of motion we easily
derive the ``Ward identities'':
$$\eqalign{
{d\over dt} \bra{G} T[y_r^i (t) y_s^j (0)]\ket{G} &= \bra{G}
T[(2p_r^i (t) - p_{r+1}^i(t) - p_{r-1}^i (t)) y_s^j (0)]\ket{G}\cr
{d\over dt} \bra{G} T[y_r^i (t) p_s^j (0)]\ket{G} &= i \delta_{ij}
\delta_{rs}\delta{(t)} + \bra{G} T [(2p_r^i (t) - p_{r+1}^i (t)
- p_{r-1}^i(t))p_s(0)]\ket{G}\cr}$$
Taking Fourier transforms of the Ward identities,
defining $k_{rs}^2 = 2\delta_{rs} -
\delta_{rs+1} - \delta_{rs-1}$, leads to
$$-i\omega {\cal E}^{ij} (\omega) = k^2 {\cal F}_{py}^{ij}
(\omega)\(wi1)$$
$$-i\omega {\cal F}_{yp}^{ij} (\omega) = \delta_{ij} I + k^2 {\cal
G}^{ij} (\omega)\(wi2)$$
$$i\omega {\cal F}_{py}^{ij} (\omega) = \delta_{ij} I + {\cal G}^{ij}
(\omega) k^2\(wi3)$$
$$i\omega {\cal E}^{ij} (\omega) = {\cal F}_{yp}^{ij} (\omega)
k^2.\(wi4)$$
The bond labels $r,s$ have been suppressed in writing these
equations, \ie\ each of the quantities
${\cal E}$, ${\cal F}_{py}$, and ${\cal G}$ is an $(M-1)\times(M-1)$
matrix, and the multiplications indicated in these
equations are matrix multiplications.
Corresponding to ${\cal E}$, ${\cal F}_{py}$, and ${\cal G}$, we shall
introduce ``irreducible parts''
$\Pi_{yy},~\Pi_{py},~\Pi_{yp},~\Pi_{pp}$. Referring to the Feynman
diagram expansion of Section 2, these quantities are defined
as the sums of all diagrams, contributing to the
respective correlation function, which cannot be cut in two by
cutting a single dashed line. See \Fig{rules}. Then we can write
an equation for each of the quantities
${\cal E}$, ${\cal F}_{py}$, and ${\cal G}$ as follows:
$${\cal G}^{ik} (\omega) = \Pi_{pp}^{ij}(\omega) [\delta_{jk} + (k^2-2)
{\cal G}^{jk}]\(pipp)$$
$${\cal F}_{yp}^{ik} (\omega) = \Pi_{yp}^{ij} (\omega) [\delta_{jk} +
(k^2-2) {\cal G}^{jk}]\(piyp)$$
$${\cal F}_{py}^{ik} (\omega) = [\delta_{ij} + {\cal G}^{ij}
(k^2-2)]\Pi_{py}^{jk}\(pipy)$$
$${\cal E}^{ik} (\omega) = \Pi_{yy}^{ik} + {\cal F}_{yp}^{ij} (k^2-2)
\Pi_{py}^{jk}.\(piyy)$$
This set of equations gives a purely algebraic determination of
the irreducible parts in terms of the full correlation functions.

Next we wish to use the Ward Identities to derive relationships
among the various $\Pi$'s.
First use \eq{wi2} and \eq{piyp} to eliminate ${\cal F}_{yp}$:
$$\delta_{ij} I + k^2 {\cal G}^{ij} (\omega) = -i\omega \Pi_{yp}^{ik}
(\omega) [\delta_{kj} + (k^2-2) {\cal G}^{kj}],\(temp)$$
and then \eq{pipp} can be used to rewrite the l.h.s. of \eq{temp}:
$$\eqalign{\delta_{ij} I + k^2 {\cal G}^{ij} (\omega)
&=\delta_{ij} I + (k^2-2) {\cal G}^{ij}
+ 2 \Pi_{pp}^{ik} [\delta_{kj} + (k^2-2) {\cal G}^{kj}]\cr
&= [\delta_{ik} + 2 \Pi_{pp}^{ik}] (\delta_{kj} + (k^2-2) {\cal
G}^{kj}).\cr}$$
Putting this into \eq{temp} gives,
$$[\delta_{ik} + 2 \Pi_{pp}^{ik} + i\omega
\Pi_{yp}^{ik}] (\delta_{kj} + (k^2-2) {\cal G}^{kj}) = 0.$$
Now {\it provided} $\delta_{kj} + (k^2-2) {\cal G}^{kj}$ is invertible,
we obtain our first relation:
$$-i\omega \Pi_{yp}^{ij} = \delta_{ij}
+ 2\Pi_{pp}^{ij}.\(temp1)$$
Next, the identities ${\cal G} (-\omega) = {\cal G}^T (\omega)$ and
${\cal F}_{py}(\omega)={\cal F}_{yp}^T (-\omega)$  yield
the transposed relationship:
$$i\omega \Pi_{py}^{ij} (\omega) = \delta_{ij} I + 2 \Pi_{pp}^{ij}
(\omega).\(temp2)$$

The last relation we seek is between $\Pi_{yy}$
and $\Pi_{pp}$. First insert \eq{wi4} in \eq{piyy} to
obtain
$${\cal E}^{ik} (\omega) = \Pi_{yy}^{ik} + i\omega {\cal E}^{ij}
(\omega) {k^2-2\over k^2} \Pi_{py}^{jk} (\omega).$$
Then use \eq{temp2} to elimintate $\Pi_{py}$,
$${\cal E}^{ik} (\omega)=\Pi_{yy}^{ik}
+ {\cal E}^{ij} (\omega) {k^2-2\over k^2}
(\delta_{jk} I + 2 \Pi_{pp}^{jk}(\omega)),$$
Leading to
$$\Pi_{yy}^{ik} (\omega)
= {\cal E}^{ij} (\omega) [{2\over k^2}
\delta_{jk} I - {k^2-2\over k^2} 2\Pi_{pp}^{jk} (\omega)].\(temp3)$$
\eq{wi2} and \eq{wi4} together give
$$\omega^2{\cal E}^{ij} {1\over k^2}
= \delta_{ij} I + k^2 {\cal G}^{ij}(\omega),$$
which, inserted into \eq{temp3}, yields
$$\eqalign{
\omega^2 \Pi_{yy}^{ik} (\omega) &= 2[\delta_{ij} I + k^2
{\cal G}^{ij} (\omega)][\delta_{jk} I - (k^2-2) \Pi_{pp}^{jk}
(\omega)]\cr
&= 2\delta_{ik} I + 2k^2 {\cal G}^{ik} (\omega) - 2 (k^2-2)
\Pi_{pp}^{ik} (\omega) - 2k^2 {\cal G}^{ij} (\omega)
(k^2-2)\Pi_{pp}^{jk}.\cr}\(yypp)$$
Now  putting $(\omega \rightarrow -\omega)$ and  taking the
transpose of  \eq{pipp} shows that
$${\cal G}^{ik} - \Pi_{pp}^{ik} = {\cal G}^{ij} (k^2-2) \Pi_{pp}^{jk}.$$
Using this result in the last term of \eq{yypp} gives,
after some
simplification, the last desired relationship:
$$\omega^2 \Pi_{yy}^{ik} (\omega) = 2\delta_{ik} I + 4 \Pi_{pp}^{ik}
(\omega),$$
or, rearranging a bit,
$$\Pi_{pp}^{ik} (\omega) = - {1\over 2} \delta_{ik} I +
{\omega^2\over 4} \Pi_{yy}^{ik} (\omega).\(finalwi)$$
The implications of the Ward identities on the irreducible
parts are summarized in \eq{temp1}, \eq{temp2},
\eq{finalwi}.

Armed with these results, we now turn to
their implications for the low lying excitations of the
polymer system. This information is given by the
location of poles in
${\cal G}$, or by the zero eigenvalues of its inverse,
which, for $\omega\approx0$, is proportional to
$$\eqalign{
\delta_{jk} - \Pi_{jk}^{pp} (\omega) (k^2-2) &= \delta_{jk} +
{1\over 2} (k^2-2) \delta_{jk} - {\omega^2\over 4} \Pi_{yy}^{jk}
(k^2-2)\cr
&={1\over 2} k^2 \delta_{jk} - {\omega^2\over 4} \Pi_{yy}^{jk} (\omega)
(k^2-2).\cr}\(inverseprop)$$
The matrix $k^2$ has eigenvalues of $O({1/M^2})$ as $M \rightarrow
\infty$, so there are eigenfrequencies $\omega^2$ of the same order.
For these we may neglect $k^2$ in the second term of
\eq{inverseprop} and evaluate
$\Pi_{yy}$ at $\omega = 0$. We therefore must search for zero
eigenvalues of the matrix
$$\omega^2 \Pi_{yy}^{ik} (0) + k^2 \delta_{ik}\(lowfreq1)$$
as $M\rightarrow \infty$. It is interesting and useful to note
that by virtue of the Ward identities, ${\cal E}^{ik}(0)
=\Pi^{ik}_{yy}(0)$. Thus there is no {\it a priori}
requirement to separate out the irreducible part: at
zero frequency the reducible parts vanish identically. Finally,
by rotational invariance, we can write
$$\Pi^{ik}_{rs}(\omega)=\delta_{ik}\Pi_{rs}(\omega), $$
so that \eq{lowfreq} can be simplified to
$$\omega^2 \Pi_{yy}(0) + k^2\(lowfreq)$$

To examine the large $M$ dependence of \eq{inverseprop} go to normal modes:
$$\eqalign{\Pi_{mn} &= {2\over M} \sum_{r,s} \sin {m\pi r\over M}
\Pi_{rs} \sin {n\pi s\over M}\cr
k_{mn}^2 &= \delta_{mn} 2 (1-\cos{n\pi\over M}),\cr}$$
after which the r.h.s. of \eq{inverseprop} becomes
$$\delta_{jk} \delta_{mn}{(1-\cos {n\pi\over M})} +
{\omega^2\over2} \Pi_{mn}(\omega)\delta_{jk}\cos{n\pi\over M}.$$
To get the spectrum of the relativistic string as
 $M\rightarrow \infty$, we should find that $\omega\propto n/M$,
which requires that $\Pi_{mn}(0)$ must be
diagonal and independent of $m$ for small ${m/M}$ or ${n/M}$.  The condition
for this
to be true is that, away from $r=2, M$,
$\Pi_{rs}\rightarrow0$ sufficiently rapidly as $\mid r-s\mid \rightarrow
\infty$.

To see this, calculate
$$\Pi_{mn} = {2\over M} \sum_{r,s} \Pi_{r+s,s} \sin
{m\pi(r+s)\over M} \sin {n\pi s\over M}.
$$
For $r,s$ away from the endpoints, $\Pi_{rs}$ should be only a function
of $r-s$, so denoting it as $\Pi({r-s})$, we have
$$\eqalign{
\Pi_{mn}
&={2\over M}\sum_{s,r}\Pi(r)
\sin {n\pi s\over M} (\sin {m\pi s\over M}
\cos {m\pi r\over M} + \cos {m\pi s\over M} \sin {m\pi r\over M})\cr
&= \delta_{mn} \Bigl( \sum_r \cos {m\pi r\over M} \Pi(r)\Bigr) +
\Bigl(\sum_r \Pi(r)
\sin {m\pi r\over M}\Bigr) {2\over M} \sum_s
\sin {n\pi s\over M} \cos {m\pi s\over M}\cr
&\approx\delta_{mn} [\sum_r\Pi(r)]\cr}$$
as $M\rightarrow\infty$, provided $\Pi^{jk}(r)$
vanishes rapidly enough as $r\rightarrow\infty$.
This is certainly true to any finite order of perturbation
theory.
Making this assumption nonperturbatively, we have,
$$\omega^2 \sum_r \Pi(r) \approx -2 (1-\cos
{m\pi\over M})= -4\sin^2 {m\pi\over 2M}\approx
-{m^2\pi^2\over (M)^2}.$$
Since $\omega$ is to be identified with
$\epsilon P^-$, and $P^+=\epsilon M$,
the (mass)$^2=2P^+P^-$ is given by $2m\pi/\sqrt{-\sum_r\Pi(r)}$,
by definition, the coefficient of $m$ in this formula is $1/\alpha^\prime
=2\pi T_0$:
$$T_0={B\over\sqrt{-\sum_r\Pi(r)}}
\qquad\qquad\alpha^\prime={1\over2\pi B}\sqrt{-\sum_r\Pi(r)}
\(exacttension)
$$
\chapter{Random Phase Approximation and First Corrections}

The Random Phase Approximation (RPA) is to calculate $\Pi_{rs}^{ij}$ replacing
$H$ by $H_0$:
$$\Pi_{{\rm RPA},rs}^{ij}(0) = \delta_{rs} \Pi^{ij} (0)$$
where
$$\Pi^{ij}(0) = - i \int_{-\infty}^\infty dt \langle0\mid T [y^i (t) y^j
(0)]\mid 0\rangle$$
corresponding to the one body problem
$$H_{1{\rm body}} = {\bf p}^2 -
{\cal V}({\bf y})$$

Putting in a complete set of states:
$$\eqalign{
\Pi^{ij} (0) &= -i\int_{-\infty}^\infty dt [\theta(t)\sum_{n\not=
0} {\langle 0\mid
y^i (0) \mid n\rangle\langle n\mid y^j (0)}\ket{0} e^{-i(E_n-E_0)t} +\cr
&\quad\quad\quad\quad\quad
\theta(-t) \sum_{n\not= 0} \langle0\mid y^j (0)\mid n\rangle
\langle n\mid y^i (0)\mid 0\rangle e^{+i(E_n-E_0)t}\cr
&= -2 \sum_{n\neq0} {\langle0\mid {\bf y}\mid n\rangle\cdot
\langle n\mid {\bf y}\mid 0\rangle\over
E_n - E_0} \left({\delta_{ij}\over 2}\right)\cr}$$
$$\omega_m^{{\rm RPA}} \left\{ \sum_{n\not= 0} {{\bf y}_{0n} \cdot
{\bf y}_{n0}\over E_n - E_0}
\right\}^{1/2} \approx {m\pi\over M}$$
{} From which, the Regge slope is
$$\alpha_{{\rm RPA}}^\prime = {1\over 2\pi}
\left\{ \sum_{m\not= 0} {{\bf y}_{0n}\cdot
{\bf y}_{n0}\over E_n - E_0}\right\}^{1/2},$$
a result first obtained by Goldstone\[goldstonepriv].

On the face of it, it would seem that the validity of RPA
depends on the smallness of the perturbation. That this
is not necessarily so is shown by the fact that
for $-{\cal V}=K{\bf y}^2$ the RPA gives the exact
result for the string tension, even though the ground
state energy receives finite corrections. This is by no
means transparent in terms of diagrams: the corrections
to the string tension cancel, apparently miraculously.
One can expect that potentials that are close to
quadratic near the minimum will lead to a quite
accurate RPA.

However, probably the most interesting
candidate string bit potential is a local
one, ${\cal V}({\bf y})=\lambda_0
\delta({\bf y})$. In two transverse dimensions
(corresponding to four dimensional space-time!), this
is a na\"ively scale invariant potential, although
ultraviolet divergences give logarithmic scaling
violations. Scale invariance is a desirable
property of the microscopic potential
since, as Gross and Mende\[grossmende]\ and others have emphasized,
it should be enjoyed by string theory
at short distances. Even apart from
its appeal for string bit dynamics,
this potential is drastically
different from a quadratic one, which makes it
a good test case for assessing the accuracy of
RPA.
Thus in this section we consider corrections
to RPA for this special potential.

We begin by reviewing the basic features of the one body
problem represented by
$$h_{1{\rm body}}={\bf p}^2-\lambda_0\delta({\bf y}),\(1body)
$$
as it determines the zeroth order in time dependent perturbation
theory. First consider the bound state problem. For an energy
eigenvalue $E=-B$, $B>0$, the momentum space wave function
is easily shown to be given by
$$\phi({\bf p})={\lambda_0\psi({\bf 0})\over{\bf p}^2+B}.$$
Since $\psi({\bf 0})=\int d{\bf p}\phi({\bf p})/(2\pi)^2$,
the quantization condition for $B$ is obtained by integrating both
sides over ${\bf p}$:
$$1=\lambda_0\int {d{\bf p}\over 4\pi^2}{1\over{\bf p}^2+B}.$$
In two dimensions the integral diverges in the ultraviolet,
which requires asymptotic freedom ($\lambda_0\rightarrow0$)
in order for $B$ to be finite. With an ultraviolet cutoff
$\Lambda$ on the momentum integral, we find
$$B=\Lambda^2 e^{-4\pi/\lambda_0},\(transmute)$$
that is we have dimensional transmutation\[thornweeparton], in
which the dimensionless $\lambda_0$ is traded for the
dimensionful $B$. In the limit $\Lambda\rightarrow\infty$,
$\psi({\bf 0})$ is infinite but $\lambda_0\psi({\bf 0})$
is finite. The normalized wave function is
$$\eqalign{
\psi({\bf y})\equiv R_B(|{\bf y}|)
=&\sqrt{B\over\pi}\int {d{\bf p}\over 2\pi}
{e^{i{\bf p}\cdot{\bf y}}\over{\bf p}^2+B}\cr
=&\sqrt{B\over\pi}K_0(|{\bf y}|\sqrt{B}),\cr}
\(ground1)$$
where $K_0$ is the zero order modified Bessel
function of the third kind.

Clearly this zero angular momentum state is the only
discrete state. The rest of the $s$-wave spectrum is
positive energy continuum. For each energy $E\equiv k^2$,
there is one state whose wave function is that unique
linear combination of Bessel functions
$J_0(k|{\bf y}|)$ and
$Y_0(k|{\bf y}|)=N_0(k|{\bf y}|)$ which is
orthogonal to \eq{ground1}:
$$R_k(|{\bf y}|)={1\over\sqrt{1+\pi^2/\ln^2(B/k^2)}}
\left[J_0(k|{\bf y}|)+{\pi\over\ln(B/k^2)}Y_0(k|{\bf y}|)\right].
\(scont1)
$$
where we have normalized $R_k$ according to
$$\int d{\bf y}R_k(|{\bf y}|)R_{k^\prime}(|{\bf y}|)
={2\pi\over k}\delta(k-k^\prime).\(norm)$$
The radial wave functions of the states with
nonzero angular momentum $L_z=m$ vanish at the origin and
so coincide with the corresponding free wave functions,
namely $R_k^m(|{\bf y}|)=J_m(k|{\bf y}|)$.

{} From the above information we can easily construct the
Green function
$$G_E({\bf y},{\bf z})\equiv
\bra{{\bf y}}{1\over h_{1{\rm body}}-E-i\epsilon}\ket{{\bf z}}.$$
The $s$-wave contribution to $G_E$ is just
$$G^{s-{\rm wave}}_E({\bf y},{\bf z})
={R_B(|{\bf y}|)R_B(|{\bf z}|)\over-B-E-i\epsilon}
+\int_0^\infty{kdk\over2\pi}
{R_k(|{\bf y}|)R_k(|{\bf z}|)\over k^2-E-i\epsilon}.
$$
The non $s$-wave contribution is just the free
Green function minus its $s$-wave. Thus we have
$$\eqalign{
G_E({\bf y},{\bf z})=&\int{d{\bf p}\over(2\pi)^2}
{e^{i{\bf p}\cdot({\bf y}-{\bf z})}\over {\bf p}^2-E-i\epsilon}
+{R_B(|{\bf y}|)R_B(|{\bf z}|)\over-B-E-i\epsilon}\cr
&\qquad+\int_0^\infty{kdk\over2\pi}
{R_k(|{\bf y}|)R_k(|{\bf z}|)-J_0(k|{\bf y}|)
J_0(k|{\bf z}|)\over k^2-E-i\epsilon}\cr}
\(green1).
$$
By working directly with the two dimensional
partial differential equation that determines $G_E$, it is
straightforward to show that an alternate
representation is
$$\eqalign{
G_E({\bf y},{\bf z})=&\int{d{\bf p}\over(2\pi)^2}
{e^{i{\bf p}\cdot({\bf y}-{\bf z})}\over {\bf p}^2-E-i\epsilon}\cr
&\qquad+{1\over\pi\ln(-(E+i\epsilon)/B)}K_0(|{\bf y}|\sqrt{-E-i\epsilon})
K_0(|{\bf z}|\sqrt{-E-i\epsilon}),\cr}
\(green1alt)
$$
Apart from the distribution of $i\epsilon$'s, $G_{\omega}$ is just
the Fourier transform of the propagator \eq{propagator}\
needed for developing time dependent perturbation theory. To
get the right boundary conditions for the propagator, one
simply has to change the $-i\epsilon$ in the middle term
of \eq{green1} to a $+i\epsilon$.
To keep track of this change, it is convenient
to define a reduced Green function $\tilde G_E$ which doesn't
have the bound state pole pole:
$${\tilde G_E}\equiv G_E-{R_B(|{\bf y}|)
R_B(|{\bf z}|)\over-B-E-i\epsilon}.
$$
Then the propagator \eq{propagator}
can be expressed as
$$\eqalign{
\bra{0}T[\psi_r^I({\bf y}, t)\psi_s^{I\dagger}({\bf z},0)]\ket{0}
=&-i\delta_{rs}\int{d\omega\over 2\pi}e^{-i\omega t}
\left[{\tilde G_{\omega}}+
{R_B(|{\bf y}|)R_B(|{\bf z}|)\over-B-\omega+i\epsilon}\right]\cr
\equiv&-i\delta_{rs}\int{d\omega\over 2\pi}e^{-i\omega t}
{\hat G_{\omega}}\cr}
\(propagatordel)
$$
With this explicit representation for the propagator we are now
in a position to write down and evaluate the expressions
assigned to the first few diagrams in time dependent
perturbation theory.

Focusing on the string tension, we have seen in the previous
section that we need to compute $\Pi_{yy}$ at zero frequency.
To lowest order in perturbation theory, we must evaluate
only the first diagram of \Fig{rpaapprox}. (The remaining
diagrams in \Fig{rpaapprox} are reducible, and in any case
vanish at zero frequency.) In next order we
have the diagrams of \Fig{corrections}. To this order
irreducible diagrams with the external sources attached
to different lines vanish by rotational invariance.

\insertfigure{corrections}
{\centerline{\psfig{figure=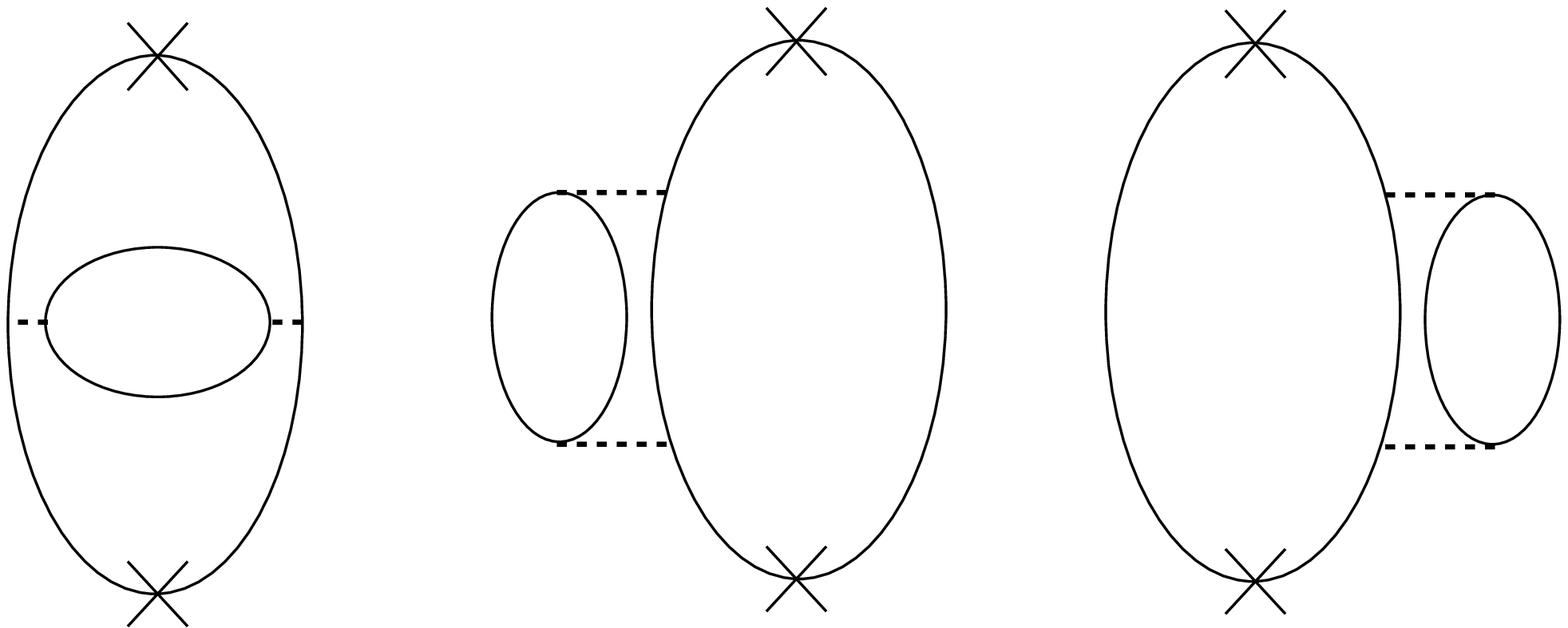,height=4cm}}}
{First order corrections to RPA.}{\vbox{\vskip4cm}}

Turning first to the lowest order diagram, the Feynman rules
assign to this diagram the expression
$$\eqalign{
&(-i)(-i)^2(-1)\delta_{rs}\int {d\omega\over2\pi}
\int d{\bf y}d{\bf z}\cr
&\quad y^j\left({\tilde G_{\omega}}({\bf y},{\bf z})
+{R_B({\bf y})R_B({\bf z})\over-B-\omega+i\epsilon}\right)
z^k\left({\tilde G}_{\omega}({\bf z},{\bf y})
+{R_B({\bf z})R_B({\bf y})\over-B-\omega+i\epsilon}\right).\cr}
$$
The prefactor phases have the following origins: the first
$-i$ from the definition of $\Pi$, the $(-i)^2$ from the
two propagators, and the $(-1)$ from the closed fermion loop.
The $\omega$ integral can immediately be done by closing the
contour in the upper half $\omega$ plane, picking up the
pole at $\omega=-B+i\epsilon$. (The singularities from $\tilde G$
are all in the lower half plane.) The result is
$$-2\delta_{rs}\int d{\bf y}d{\bf z}R_B({\bf y})y^j{\tilde G}_{-B}({\bf y},
{\bf z})z^kR_B({\bf z}),$$
which is identical to the result of the previous section.
Because $R_B$ is an $s$-wave state, the presence of $y^j$ and
$z^k$ cause all $s$-wave contributions to $\tilde G$ to
decouple, so
the latter can be replaced with the free Green function.
Working in momentum space we get
$$-2\delta_{rs}{\delta_{jk}\over2}{B\over\pi}\int d{\bf p}
{4{\bf p}^2\over({\bf p}^2+B)^5}=-4B\delta_{rs}\delta_{jk}
\int_B^\infty{v-B\over v^5}dv=-{1\over3B^2}\delta_{rs}\delta_{jk}.
$$
This approximation for $\Pi_{yy}$ is, as we have discussed,
the random phase approximation, giving a string tension of
$T_0\equiv1/2\pi\alpha^\prime\approx T_0^{RPA}= B\sqrt{3}$.

Corrections to RPA are given by the three diagrams in
\Fig{corrections}. The first
 diagram represents the expression
$$\eqalign{
&-2\delta_{rs}i(-i)^6(-1)^2i^2\int d{\bf y}d{\bf z}d{\bf u}d{\bf v}
d{\bf x}d{\bf w}
\int {d\omega_1\over2\pi}{d\omega_2\over2\pi}{d\omega_3\over2\pi}\cr
&y^j{\hat G}_{\omega_2}({\bf y},{\bf u})(-i\nabla_u^m)
{\hat G}_{\omega_1}({\bf u},{\bf z})z^k
{\hat G}_{\omega_1}({\bf z},{\bf v})(-i\nabla_v^n)
{\hat G}_{\omega_2}({\bf v},{\bf y})\cr
&(-i\nabla_x^n){\hat G}_{\omega_1-\omega_2+\omega_3}({\bf x},{\bf w})
(-i\nabla_w^m){\hat G}_{\omega_3}({\bf w},{\bf x})\cr}
$$
The factor of two comes from the two possible terms in
the vertex function \eq{vertex}; this assumes that
$r,s$ are well away from 2 and $M$.
To begin reducing this expression first note that the $\omega_3$
integral can be closed in the upper half plane where there is
only a simple pole at $\omega_3=-B+i\epsilon$. Thus
$$\eqalign{
&\int {d\omega_3\over2\pi}\int d{\bf x}d{\bf w}
(-i\nabla_x^n){\hat G}_{\omega_1-\omega_2+\omega_3}({\bf x},{\bf w})
(-i\nabla_w^m){\hat G}_{\omega_3}({\bf w},{\bf x})=\cr
&-i\int d{\bf x}d{\bf w}[R_B({\bf x})(-i\nabla_x^n)
{\tilde G}_{\omega_1-\omega_2-B}({\bf x},{\bf w})
(-i\nabla_w^m)R_B({\bf w})+\cr
&R_B({\bf w})(-i\nabla_w^m)
{\tilde G}_{\omega_2-\omega_1-B}({\bf w},{\bf x})
(-i\nabla_x^n)R_B({\bf x})]\cr
=&-i{B\over\pi}\int d{\bf p}{p^mp^n\over({\bf p}^2+B)^2}
\left[{1\over{\bf p}^2+B-\omega_1+\omega_2-2i\epsilon}+
{1\over{\bf p}^2+B-\omega_2+\omega_1-2i\epsilon}\right]\cr
=&{-iB\delta_{mn}\over2}\int_B^\infty dt{t-B\over t^2}
\left[{1\over t-\omega_1+\omega_2-2i\epsilon}+
{1\over t-\omega_2+\omega_1-2i\epsilon}\right].\cr}
\(bubble)
$$
This same integral appears in the other two diagrams, so we
shall use this result again.

The first diagram has now been reduced to
$$\eqalign{
&-2i\delta_{rs}\int d{\bf y}d{\bf z}d{\bf u}d{\bf v}
\int {d\omega_1\over2\pi}{d\omega_2\over2\pi}\cr
&y^j{\hat G}_{\omega_2}({\bf y},{\bf u})(-i\nabla_u^m)
{\hat G}_{\omega_1}({\bf u},{\bf z})z^k
{\hat G}_{\omega_1}({\bf z},{\bf v})(-i\nabla_v^n)
{\hat G}_{\omega_2}({\bf v},{\bf y})\cr
&{-iB\delta_{mn}\over2}\int_B^\infty dt{t-B\over t^2}
\left[{1\over t-\omega_1+\omega_2-2i\epsilon}+
{1\over t-\omega_2+\omega_1-2i\epsilon}\right].\cr}
\(1diagram)
$$
Now consider separately the two terms in the last factor. In
the first term close the $\omega_1$ integral in the upper
half plane, but in the second close the $\omega_2$ integral
in the upper half plane. In each case there is only a
simple pole at $\omega_1=-B+i\epsilon$, respectively
$\omega_2=-B+i\epsilon$. The potential double poles at these
locations don't contribute because they correspond
to two $s$-wave states separated by a vector operator.
(This will {\it not} apply to the other two diagrams!)
After this the remaining $\omega$ integral is done by
again closing in the upper half plane. Altogether there
are four separate contributions, each of which can
be thought of as a ground state
expectation value in the one body problem. The results of
the steps described so far can be represented, in
a self-explanatory notation, as
$$\eqalign{
{-B}\delta_{rs}\int_B^\infty dt{t-B\over t^2}
[&\bra{B}y^k{1\over{\bf p}^2+B}
p^m{\hat G}_{-t-B}y^j{1\over{\bf p}^2+B+t}p^m\ket{B}\cr
&\qquad-{1\over t}\bra{B}y^k{1\over{\bf p}^2+B}
p^m\ket{B}\bra{B}y^j{1\over{\bf p}^2+B}p^m\ket{B}\cr
&+\bra{B}p^m{1\over{\bf p}^2+B+t}
y^j{\hat G}_{-t-B}p^m{1\over{\bf p}^2+B}y^k\ket{B}\cr
&\qquad-{1\over t}\bra{B}p^m{1\over{\bf p}^2+B}
y^j\ket{B}\bra{B}p^m{1\over{\bf p}^2+B}y^k\ket{B}\cr
&+\bra{B}p^m{1\over{\bf p}^2+B+t}
y^k{\hat G}_{-t-B}p^m{1\over{\bf p}^2+B}y^j\ket{B}\cr
&\qquad-{1\over t}\bra{B}p^m{1\over{\bf p}^2+B}
y^k\ket{B}\bra{B}p^m{1\over{\bf p}^2+B}y^j\ket{B}\cr
&+\bra{B}y^j{1\over{\bf p}^2+B}
p^m{\hat G}_{-t-B}y^k{1\over{\bf p}^2+B+t}p^m\ket{B}]\cr
&\qquad-{1\over t}\bra{B}y^j{1\over{\bf p}^2+B}
p^m\ket{B}\bra{B}y^k{1\over{\bf p}^2+B}p^m\ket{B}]\cr}
\(1diagramred)
$$
where we have replaced the first and last Green functions in
each term by the free ones, since the $s$-wave parts decouple
in these locations. We postpone explicit evaluation until we
have reduced the remaining diagrams. In addition, $\epsilon$
may be safely set to zero in all of the terms in \eq{1diagramred}.

We present the expressions for
the second and third diagrams of \Fig{corrections}
after the first simplification step analogous to
\eq{1diagram}.  The second diagram leads to
$$\eqalign{
&-2i\delta_{rs}\int d{\bf y}d{\bf z}d{\bf u}d{\bf v}
\int {d\omega_1\over2\pi}{d\omega_2\over2\pi}\cr
&y^j{\hat G}_{\omega_1}({\bf y},{\bf z})z^k
{\hat G}_{\omega_1}({\bf z},{\bf u})
(-i\nabla_u^m){\hat G}_{\omega_2}({\bf u},{\bf v})(-i\nabla_v^m)
{\hat G}_{\omega_1}({\bf v},{\bf y})\cr
&{-iB\over2}\int_B^\infty dt{t-B\over t^2}
\left[{1\over t-\omega_1+\omega_2-2i\epsilon}+
{1\over t-\omega_2+\omega_1-2i\epsilon}\right],\cr}
\(2diagram)
$$
and the third to
$$\eqalign{
&-2i\delta_{rs}\int d{\bf y}d{\bf z}d{\bf u}d{\bf v}
\int {d\omega_1\over2\pi}{d\omega_2\over2\pi}\cr
&y^j{\hat G}_{\omega_1}({\bf y},{\bf u})(-i\nabla_u^m)
{\hat G}_{\omega_2}({\bf u},{\bf v})
(-i\nabla_v^m){\hat G}_{\omega_1}({\bf v},{\bf z})
z^k{\hat G}_{\omega_1}({\bf z},{\bf y})\cr
&{-iB\over2}\int_B^\infty dt{t-B\over t^2}
\left[{1\over t-\omega_1+\omega_2-2i\epsilon}+
{1\over t-\omega_2+\omega_1-2i\epsilon}\right].\cr}
\(3diagram)
$$
The next step is to consider separately the expressions
arising from the two terms in the last factor. For the
first term, close the $\omega_1$ contour up and for the
second close the $\omega_2$ contour up. Here the asymmetry
between $\omega_1$ and $\omega_2$ causes some differences
{} from the corresponding simplification of the first
diagram. The first terms involve both a double pole and
and three simple
poles at $\omega_1=-B+i\epsilon$. Whereas the second terms
involve only a single simple pole. The four simple pole
contributions from each diagram lead to expressions very
like \eq{1diagramred}. Namely,
$$\eqalign{
{-B}\delta_{rs}\int_B^\infty dt{t-B\over t^2}
[&\bra{B}y^k{1\over{\bf p}^2+B}
p^m{\hat G}_{-t-B}p^m{1\over{\bf p}^2+B}y^j\ket{B}\cr
&\qquad-{1\over t}\bra{B}y^k{1\over{\bf p}^2+B}
p^m\ket{B}\bra{B}p^m{1\over{\bf p}^2+B}y^j\ket{B}\crr
&+\bra{B}p^m{1\over{\bf p}^2+B+t}
p^m{\tilde G}_{-B}y^j{1\over{\bf p}^2+B}y^k\ket{B}\crr
&+\bra{B}p^m{1\over{\bf p}^2+B+t}
y^j{\hat G}_{-t-B}y^k{1\over{\bf p}^2+B+t}p^m\ket{B}\crr
&\qquad-{1\over t}\bra{B}p^m{1\over{\bf p}^2+B}
y^j\ket{B}\bra{B}y^k{1\over{\bf p}^2+B}p^m\ket{B}\crr
&+\bra{B}y^j{1\over{\bf p}^2+B}
y^k{\tilde G}_{-B}p^m{1\over{\bf p}^2+B+t}p^m\ket{B}]\cr
}
\(2diagramsimp)
$$
for the first diagram, and
$$\eqalign{
{-B}\delta_{rs}\int_B^\infty dt{t-B\over t^2}
[&\bra{B}y^k{1\over{\bf p}^2+B}
y^j{\tilde G}_{-B}p^m{1\over{\bf p}^2+B+t}p^m\ket{B}\crr
&+\bra{B}p^m{1\over{\bf p}^2+B+t}
p^m{\tilde G}_{-B}y^k{1\over{\bf p}^2+B}y^j\ket{B}\crr
&+\bra{B}p^m{1\over{\bf p}^2+B+t}
y^k{\hat G}_{-t-B}y^j{1\over{\bf p}^2+B+t}p^m\ket{B}\crr
&\qquad-{1\over t}\bra{B}p^m{1\over{\bf p}^2+B}
y^k\ket{B}\bra{B}y^j{1\over{\bf p}^2+B}p^m\ket{B}\crr
&+\bra{B}y^j{1\over{\bf p}^2+B}
p^m{\hat G}_{-t-B}p^m{1\over{\bf p}^2+B}y^k\ket{B}]\crr
&\qquad-{1\over t}\bra{B}y^j{1\over{\bf p}^2+B}
p^m\ket{B}\bra{B}p^m{1\over{\bf p}^2+B}y^k\ket{B}]\crr
}
\(3diagramsimp)
$$
for the second. \eq{2diagramsimp} and \eq{3diagramsimp} do
not include the double pole contribution to the integral
of the first terms over $\omega_1$. These arise from the
ground state contributions to the second and fourth ${\hat G}$'s
in \eq{2diagram} and to the first and third ${\hat G}$'s in
\eq{3diagram}. Evaluating the residue of the double pole in
$\omega_1$ and then doing the $\omega_2$ integral leads to
the double pole contributions
$$\eqalign{
&B\delta_{rs}\int_B^\infty dt{t-B\over t^2}
{d\over d\omega}\left[\bra{B}y^j{1\over{\bf p}^2+B-\omega}
y^k\ket{B}\bra{B}p^m{1\over{\bf p}^2+B+t-\omega}
p^m\ket{B}\right]_{\omega=0}\cr
&\qquad=B\delta_{rs}\int_B^\infty dt{t-B\over t^2}
[\bra{B}y^j{1\over({\bf p}^2+B)^2}
y^k\ket{B}\bra{B}p^m{1\over{\bf p}^2+B+t}p^m\ket{B}\cr
&\qquad\qquad+\bra{B}y^j{1\over{\bf p}^2+B}
y^k\ket{B}\bra{B}p^m{1\over({\bf p}^2+B+t)^2}p^m\ket{B}]\cr}
\(2diagramdoub)
$$
and
$$\eqalign{
&B\delta_{rs}\int_B^\infty dt{t-B\over t^2}{d\over d\omega}\left[
\bra{B}y^k{1\over{\bf p}^2+B-\omega}
y^j\ket{B}\bra{B}p^m{1\over{\bf p}^2+B+t-\omega}
p^m\ket{B}\right]_{\omega=0}\cr
&\qquad=B\delta_{rs}\int_B^\infty dt{t-B\over t^2}
[\bra{B}y^k{1\over({\bf p}^2+B)^2}
y^j\ket{B}\bra{B}p^m{1\over{\bf p}^2+B+t}p^m\ket{B}\cr
&\qquad\qquad+\bra{B}y^k{1\over{\bf p}^2+B}
y^j\ket{B}\bra{B}p^m{1\over({\bf p}^2+B+t)^2}p^m\ket{B}]\cr}
\(3diagramdoub)
$$
respectively.

Notice that lines 2 and 5 of \eq{2diagramsimp} and lines 4 and
6 of \eq{3diagramsimp} cancel lines 2, 4, 6, and 8 of
\eq{1diagramred} by virtue of the fact that
$$\bra{B}y^j{1\over{\bf p}^2+B}p^m\ket{B}=
-\bra{B}p^m{1\over{\bf p}^2+B}y^j\ket{B}$$
by time reversal invariance. This cancellation
together with similarity of structure make it helpful to regroup
the various contributions by gathering together all terms
with the central Green function ${\hat G}_{-t-B}$ in
group 1,
$$\eqalign{
{-B}\delta_{rs}\int_B^\infty dt{t-B\over t^2}
[&\bra{B}y^k{1\over{\bf p}^2+B}
p^m{\hat G}_{-t-B}y^j{1\over{\bf p}^2+B+t}p^m\ket{B}\cr
&+\bra{B}p^m{1\over{\bf p}^2+B+t}
y^j{\hat G}_{-t-B}p^m{1\over{\bf p}^2+B}y^k\ket{B}\cr
&+\bra{B}p^m{1\over{\bf p}^2+B+t}
y^k{\hat G}_{-t-B}p^m{1\over{\bf p}^2+B}y^j\ket{B}\cr
&+\bra{B}y^j{1\over{\bf p}^2+B}
p^m{\hat G}_{-t-B}y^k{1\over{\bf p}^2+B+t}p^m\ket{B}]\cr
&+\bra{B}y^k{1\over{\bf p}^2+B}
p^m{\hat G}_{-t-B}p^m{1\over{\bf p}^2+B}y^j\ket{B}\cr
&+\bra{B}p^m{1\over{\bf p}^2+B+t}
y^j{\hat G}_{-t-B}y^k{1\over{\bf p}^2+B+t}p^m\ket{B}\cr
&+\bra{B}p^m{1\over{\bf p}^2+B+t}
y^k{\hat G}_{-t-B}y^j{1\over{\bf p}^2+B+t}p^m\ket{B}\cr
&+\bra{B}y^j{1\over{\bf p}^2+B}
p^m{\hat G}_{-t-B}p^m{1\over{\bf p}^2+B}y^k\ket{B}],\cr
}
\(group1)
$$
and all those with central Green function ${\tilde G}_{-B}$ into group
2,
$$\eqalign{
{-B}\delta_{rs}\int_B^\infty dt{t-B\over t^2}
[&\bra{B}p^m{1\over{\bf p}^2+B+t}
p^m{\tilde G}_{-B}y^j{1\over{\bf p}^2+B}y^k\ket{B}\cr
&+\bra{B}y^j{1\over{\bf p}^2+B}
y^k{\tilde G}_{-B}p^m{1\over{\bf p}^2+B+t}p^m\ket{B}\cr
&+\bra{B}y^k{1\over{\bf p}^2+B}
y^j{\tilde G}_{-B}p^m{1\over{\bf p}^2+B+t}p^m\ket{B}\cr
&+\bra{B}p^m{1\over{\bf p}^2+B+t}
p^m{\tilde G}_{-B}y^k{1\over{\bf p}^2+B}y^j\ket{B}].\cr
}
\(group2)
$$
The double pole contributions are also naturally associated with
group 2.

In the appendix we describe the detailed evaluation of all
these expressions. By far the largest contribution comes
{} from the second and third diagrams of \Fig{corrections}.
The total contribution of the first diagram is
approximately $-{0.055197/B^2}$ or about 16.6\% of the
random phase approximation. Compare this to the total contribution
of the second and third diagrams of $+{0.50913/B^2}$
or $-153\%$ of the RPA. In the next section we show that this
huge correction can be attributed to a substantial shift
in the ionization energy of the open polymer. The large terms
in perturbation theory can be identified as ``self energy''
corrections which can be easily summed as a geometric
series in the usual way.

\chapter{Summing Large Corrections}
We have found very large corrections to RPA in next to
lowest order of perturbation theory. But we have also seen
that all the large contributions come from diagrams
that have the interpretation as self energy bubbles
on the propagators entering the RPA. Thus we can hope
that an improved RPA would result if we use the
full propagators rather than the bare ones. In other words,
we should take the sum of diagrams shown in \Fig{hartree}
as the propagator to use in RPA. An even more sophisticated
improvement would be to replace each bubble in \Fig{hartree}
with the RPA sum of \Fig{rpaapprox}.

\insertfigure{hartree}
{\centerline{\psfig{figure=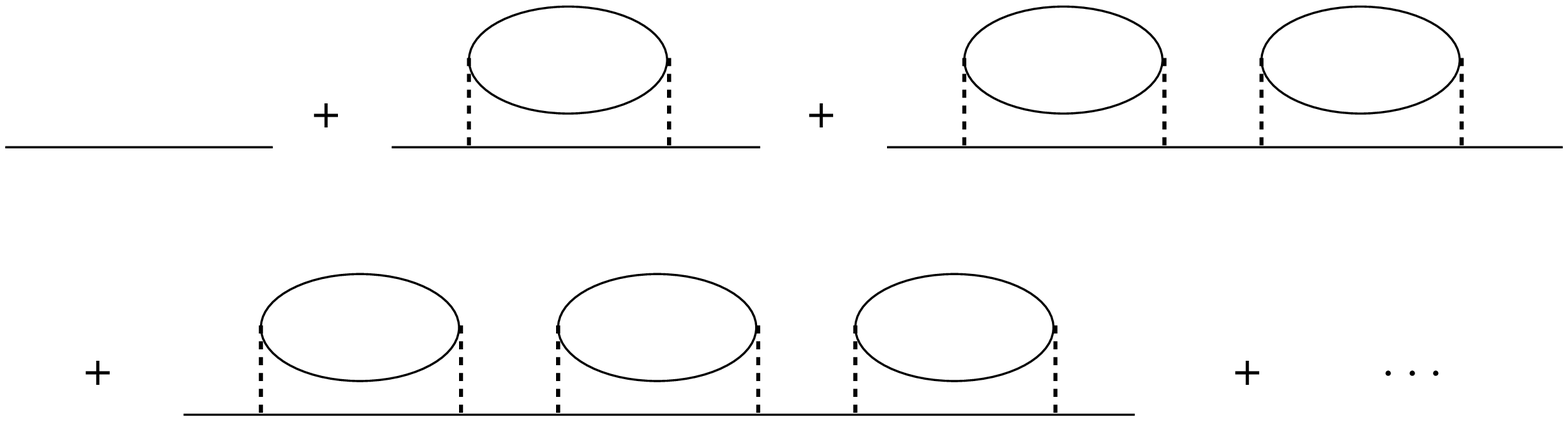,height=3.5cm}}}
{Summing self energy diagrams.}{\vbox{\vskip3.5cm}}

More generally we can define a ``self energy'' part, $\Sigma$,
so that $i\Sigma$
is the sum of all the (amputated) one particle irreducible
diagrams contributing to the correlation function of
$\psi$ and $\psi^\dagger$. Then the full propagator
$$\bra{G}T[\psi_r({\bf y},t)\psi^\dagger_s({\bf z},0)]\ket{G}
\equiv -i\delta_{r,s}\Delta$$
satisfies an integral equation
$$\Delta={\hat G} + {\hat G}\Sigma\Delta.  \(inteq1)$$
Because the operator $\psi_r$ annihilates a
bond in the polymer, the location in $-\omega$ of the
lowest pole in $\Delta$ gives the
``exact'' ionization energy, which in zeroth order of
perturbation theory is just $B$, the lowest pole in ${\hat G}$.
To find this lowest pole it is sufficient to look at a
matrix element of $\Delta$ between any two states
 which couple to it. Assuming that
the zeroth order ground state continues to couple to this
pole, it is convenient to use a new concept of
``irreducible'' to mean that one can't cut the diagram in two
by cutting a {\it single line in its ground state}.
This new irreducible part will then be called $\Sigma^\prime$.
A formula for $\Sigma^\prime$ is obtained by splitting
${\hat G}={\tilde G} + G_0$ (see \eq{propagatordel})
where, in frequency space,
$$G_0\equiv {\ket{B}\bra{B}\over-B-\omega +i\epsilon}.$$
Then we have the integral equation
$$\Sigma^\prime=\Sigma+\Sigma {\tilde G}\Sigma^\prime.
\(sigmaprimeeq) $$
That is, we put back into $\Sigma^\prime$ all those
reducible contributions for which the single line
is in an excited state.

With these definitions a little algebraic gymnastics leads
to the following formula for the full propagator:
$$\Delta={\tilde G} + {\tilde G}\Sigma^\prime {\tilde G}
+(I+{\tilde G}\Sigma^\prime)
{1\over-B-\omega-\Sigma^\prime_0+i\epsilon}P_0
(I+\Sigma^\prime {\tilde G})\(fullprop)
$$
where $P_0\equiv\ket{B}\bra{B}$ is the projector
onto the zeroth order ground state,
and $\Sigma^\prime_0\equiv\bra{B}\Sigma^\prime\ket{B}$
a real number. As we explain in the Appendix, the
first step in the evaluation of a typical diagram is
to close one or more of the frequency integrals in
the upper half plane, picking up some of the ground state
poles in the various  ${\hat G}_\omega$'s. In a more
sophisticated application of perturbation theory,
where $\Delta$ is used in place of ${\hat G}$, this step
will pick up the pole at $\omega=\omega_0$ where
$$D(\omega_0)\equiv-B-\omega_0-\Sigma^\prime_0(\omega_0)
+i\epsilon=0. \(denom)$$
The residue of this pole is
$${1\over D^\prime(\omega_0)}
(I+{\tilde G}_{\omega_0}\Sigma^\prime(\omega_0))P_0
(I+\Sigma^\prime(\omega_0) {\tilde G}_{\omega_0})
={1\over D^\prime(\omega_0)}
G_{\omega_0}\Sigma^\prime(\omega_0)P_0
\Sigma^\prime(\omega_0)G_{\omega_0}.
\(fullpropres)$$
Note that the form on the r.h.s. would be meaningless if
$\omega_0=B$ since $G_{-B}$ is infinite. But since the
shift of $\omega_0$ away from its zeroth order value is
taken into account, this is not a problem. (If by some
fluke the shift were zero, then $\Sigma^\prime_{-B}$ would
supply a zero multiplying the infinity, leaving the
r.h.s. indeterminate.)

Now we can use these results to specify our improved
version of RPA. It is obtained by replacing each of the
bare propagators in the first diagram of
\Fig{rpaapprox} with the the full propagator
\eq{fullprop}. Corrections which involve ``self energy''
insertions such as the second and third diagrams of
\Fig{corrections}, will then automatically be included in
this approximation. The first diagram of \Fig{corrections}
is a typical diagram not included. Our experience
with straight perturbation theory indicates that the
diagrams not included give roughly a 15\% correction.

Inserting \eq{fullprop} into the RPA diagram, we can carry
out the $\omega$ integral, just as with the bare propagators,
by closing the contour into the upper half plane, picking up
the lowest pole at $\omega=\omega_0$, determined by \eq{denom}.
The $+i\epsilon$ of \eq{denom} amounts to the contour prescription
which states that this lowest pole is the {\it only} singularity
picked up by this procedure. The result is
$$\Pi_{yy,rs}^{{\rm Improved},jk}(0)=
-2\delta_{rs}{\bra{B}\Sigma^\prime(\omega_0)
G_{\omega_0}y^j[{\tilde G}_{\omega_0}+{\tilde G}_{\omega_0}
\Sigma^\prime(\omega_0){\tilde G}_{\omega_0}]y^kG_{\omega_0}
\Sigma^\prime(\omega_0)\ket{B}\over1+{d\over d\omega}
\Sigma^\prime_0(\omega)\mid_{\omega=\omega_0}}.
\(improvedrpa)$$
Note that the pole contribution to $\Delta$ vanishes when
sandwiched between $y^j$ and $y^k$ by rotational invariance.
If we were given the exact result for $\Sigma^\prime$ we could
use this formula to calculate the string tension. We would
first have to find $\omega_0$ by solving \eq{denom}. Then
evaluating $\Sigma^\prime$ at $\omega_0$ and plugging into
\eq{improvedrpa} would give us the required information.

Unfortunately, $\Sigma^\prime$ is not a quantity we can
evaluate exactly, and to assess the ramifications of this procedure
we must make further approximations. A plausible
approximation
would be to calculate $\Sigma$ in perturbation
theory. Thus in lowest order we would simply take the
second diagram of \Fig{hartree}, with external propagators
removed, for $\Sigma(\omega)$. One could then try to use
this approximate form for $\Sigma$ in \eq{sigmaprimeeq},
to determine $\Sigma^\prime$. Although
inexact, this is a substantial improvement over straight
perturbation theory because, for instance, the procedure
does not violate easily proved positivity constraints
on ${\cal G}_{rs}$.
Probably the approximation most consistent with RPA would be to
use the RPA sum for the bubble in the lowest order
diagram for $\Sigma$, as we described in the
opening paragraph of this section. In the remainder
of this section we attempt a simplified version of
the first mentioned scheme. The complete implementations
of the first scheme and of the second would involve
some nontrivial technical
developments, and we leave them for future work.

To get a feeling for how these considerations
affect the numerical results of the
previous section, we close this section with the
evaluation of the string tension using the lowest order
approximation to $\Sigma$,
$$\eqalign{
\Sigma(\omega)
\approx&\cr
{B\over\pi}\int d{\bf q}&{{\bf q}^2\over({\bf q}^2+B)^2}
\left(
\sum_{m>0}{p^k\ket{m}\bra{m}p^k\over E_m+{\bf q}^2+B-\omega-3i\epsilon}
+{p^k\ket{0}\bra{0}p^k\over E_0+{\bf q}^2+B-\omega+3i\epsilon}
\right),\cr
}
$$
in an approximate version of \eq{sigmaprimeeq}. Recalling that
the large corrections found in the previous section could
be traced to the contribution of the second term of
\eq{green1alt}, it would seem to be reasonable to truncate
the ${\tilde G}$ appearing in \eq{sigmaprimeeq} to
that term.  We therefore define, in notation explained in the
appendix,
$${\tilde G}\approx {\tilde G}_{{\rm trunc}}(\omega)\equiv{1\over
B\ln(-\omega/B)}
\ket{-\omega}\bra{-\omega}-{1\over-\omega-B}\ket{B}\bra{B}.
\(greentrunc)
$$
We can then use \eq{sigmaprimeeq} to determine by
straightforward algebra the matrix elements
$\bra{B}\Sigma^\prime\ket{B}$, $\bra{-\omega}\Sigma^\prime\ket{B}$,
$\bra{B}\Sigma^\prime\ket{-\omega}$, and
$\bra{-\omega}\Sigma^\prime\ket{-\omega}$
in terms of analogous matrix elements of $\Sigma$.

Define $\gamma_\omega\equiv1/\{B\ln(-\omega/B)\}$,
$\gamma_0\equiv1/(\omega+B)$, and
$$\eqalign{
&D_1\equiv 1-\gamma_\omega\bra{-\omega}\Sigma\ket{-\omega}
-\gamma_0\bra{B}\Sigma\ket{B}+\cr
&\qquad\qquad\gamma_\omega\gamma_0[\bra{B}\Sigma\ket{B}
\bra{-\omega}\Sigma\ket{-\omega}-\bra{-\omega}\Sigma\ket{B}
\bra{B}\Sigma\ket{-\omega}].\cr}$$
Then we find
$$\eqalign{
\bra{B}\Sigma^\prime\ket{B}&\approx{\bra{B}\Sigma\ket{B}-\gamma_\omega
[\bra{B}\Sigma\ket{B}
\bra{-\omega}\Sigma\ket{-\omega}-\bra{-\omega}\Sigma\ket{B}
\bra{B}\Sigma\ket{-\omega}]\over D_1}\cr
\bra{-\omega}\Sigma^\prime\ket{B}&
\approx{\bra{-\omega}\Sigma\ket{B}\over D_1}\cr
\bra{B}\Sigma^\prime\ket{-\omega}&\approx
{\bra{B}\Sigma\ket{-\omega}\over D_1}\cr
\bra{-\omega}\Sigma^\prime\ket{-\omega}&\approx\cr
&{\bra{-\omega}\Sigma\ket{-\omega}-\gamma_0
[\bra{B}\Sigma\ket{B}
\bra{-\omega}\Sigma\ket{-\omega}-\bra{-\omega}\Sigma\ket{B}
\bra{B}\Sigma\ket{-\omega}]\over D_1}\cr
}
$$
For the matrix elements of $\Sigma$, we find, after doing
angular integrals and some minor
changes of integration variables,
$$\eqalign{
 \bra{B}\Sigma\ket{B}=&B\int_1^\infty {dv\over v^2}
\int_1^\infty {du\over u^2}{(v-1)(u-1)\over u+v-1-\omega/B}\equiv
Bf_1(\rho)\cr
 \bra{B}\Sigma\ket{-\omega}=&
B\int_1^\infty {dv\over v^2}
\int_1^\infty {du\over u}{(v-1)(u-1)\over
(u-1-\omega/B)(u+v-1-\omega/B)}\equiv Bf_2(\rho)\cr
\bra{-\omega}\Sigma\ket{B}=&\bra{B}\Sigma\ket{-\omega}=Bf_2(\rho)\cr
\bra{-\omega}\Sigma\ket{-\omega}=&B\int_1^\infty {dv\over v^2}
\int_1^\infty {du\over (u-1-\omega/B)^2}
{(v-1)(u-1)\over u+v-1-\omega/B}\equiv Bf_3(\rho)\cr
}
$$
where we have defined $\rho\equiv-1-\omega/B$. The $v$ integral is
elementary, with the results
$$\eqalign{f_1(\rho)=&\int_1^\infty{(u-1)du\over u^2}
\left[{\ln(1+u+\rho)-1\over u+\rho}+{\ln(1+u+\rho)\over
(u+\rho)^2}\right]\cr
f_2(\rho)=&\int_1^\infty{(u-1)du\over u}
\left[{\ln(1+u+\rho)-1\over (u+\rho)^2}+{\ln(1+u+\rho)\over
(u+\rho)^3}\right]\cr
f_3(\rho)=&\int_1^\infty{(u-1)du}
\left[{\ln(1+u+\rho)-1\over (u+\rho)^3}+{\ln(1+u+\rho)\over
(u+\rho)^4}\right].\cr
}
$$

Putting everything together, we express all quantities in terms of
the functions $f_1,f_2,f_3$. For $D_1$ we have
$$D_1\approx1-{f_3(\rho)\over\ln(1+\rho)}+{f_1(\rho)\over\rho}
-{f_1(\rho)f_3(\rho)-f_2(\rho)^2\over\rho\ln(1+\rho)},
$$
and for the matrix elements of $\Sigma^\prime$,
$$\eqalign{
\bra{B}\Sigma^\prime\ket{B}&\approx B{f_1(\rho)
-[f_1(\rho)f_3(\rho)-f_2(\rho)^2]/\ln(1+\rho)\over D_1}\cr
\bra{-\omega}\Sigma^\prime\ket{B}&\approx
B{f_2(\rho)\over D_1}\cr
\bra{B}\Sigma^\prime\ket{-\omega}&\approx
B{f_2(\rho)\over D_1}\cr
\bra{-\omega}\Sigma^\prime\ket{-\omega}&\approx
B{f_3(\rho)
+[f_1(\rho)f_3(\rho)-f_2(\rho)^2]/\rho\over D_1}.\cr
}
$$
Using the above approximate result for
$\bra{B}\Sigma^\prime\ket{B}=\Sigma^\prime_0$ in
\eq{denom} gives the following approximate
equation for $\rho_0=-1-\omega_0/B$:
$$\rho_0-{f_1(\rho_0)
-[f_1(\rho_0)f_3(\rho_0)-f_2(\rho_0)^2]/\ln(1+\rho_0)\over D_1(\rho_0)}
\approx0.
$$
Using a simple iterative search for solutions to this
equation leads to a value for $\rho_0$ between 0.589941
and 0.589942, so we can say that $\rho_0\approx0.590$.
This corresponds to a shift in the ionization energy
{} from $B$ to $1.590B$. The same numerical
data give $1+d\Sigma^\prime_0/d\omega\approx1.06$,
which is consistent with unity in the present discussion.
Straight perturbation theory would
lead to the result $\omega_0=-(1+f_1(0))B$. But $f_1(0)=I_{11}-I_{21}
\approx0.59086$. So, curiously enough,
the improved prediction for $\omega_0$
is barely distinguishable from the calculation of the
first two orders of perturbation theory! As an
aside, we recall that in the Hartree-Fock method which
we are using here, the relationship of the shift in
ionization energy (or single particle energy) is related to
the shift in the ground state energy per bond (or per single
particle) by a factor of 2. Thus this 60\% shift in the
ionization energy, which is rather huge, corresponds to
a more moderate 30\% shift in the ground state energy.

Having approximately determined $\omega_0$, we consider the
evaluation of \eq{improvedrpa} at this value. In the
same spirit as our approximate treatment of \eq{sigmaprimeeq},
It is reasonable to approximate the first and
last $G_{\omega_0}$'s on the r.h.s. by
$$G_{\omega_0}\approx{1\over B\ln(-\omega_0/B)}\ket{-\omega_0}
\bra{-\omega_0}.
$$
Then
$$\eqalign{
&\Pi_{yy,rs}^{{\rm Improved},jk}(0)\approx\cr
&-2\delta_{rs}
{|\bra{B}\Sigma^\prime(\omega_0)\ket{-\omega_0}|^2
\over(1+{d\over d\omega}
\Sigma^\prime_0(\omega)\mid_{\omega=\omega_0})\ln^2(1+\rho_0)}
\bra{-\omega_0}y^j[{\tilde G}_{\omega_0}+{\tilde G}_{\omega_0}
\Sigma^\prime(\omega_0){\tilde G}_{\omega_0}]y^k\ket{-\omega_0}\cr}
$$
Referring to our experience with straight perturbation theory
the second term includes contributions in next to lowest
order which are relatively small, \ie\ roughly 15\%. Thus
it is fair to drop it in the present approximation. The
resulting matrix element is easily evaluated
$$\bra{-\omega_0}y^j{\tilde G}_{\omega_0}y^k\ket{-\omega_0}
={\delta_{jk}\over6B^2(1+\rho_0)^3}
\approx0.249{\delta_{jk}\over6B^2}.$$
Thus to an expected 15\% accuracy we have
$$\Pi_{yy,rs}^{{\rm Improved},jk}(0)\approx-2\delta_{rs}
{f_2(\rho_0)^2
\over D_1(\rho_0)^2\ln^2(1+\rho_0)}
\bra{-\omega_0}y^j{\tilde G}_{\omega_0}y^k\ket{-\omega_0}
\approx-0.333{\delta_{rs}\delta_{jk}\over3B^2}.
$$
Thus our improved RPA causes predicts a reduction in $\Pi$
by  approximately a factor of 3 over the na\"ive RPA. But now we have
explicitly incorporated all of the large corrections in
next order in a partial resummation of perturbation
theory. The effects left out amount to a roughly 15\%
correction in this next order, and it is reasonable
to expect that this is a rough indication of the
accuracy of our improved RPA. The factor of 3 reduction
in $\Pi$ corresponds to a factor of $\sqrt{3}$ increase
in the rest tension leading to
$$T_0^{\rm Improved~RPA}={1\over\alpha^{\prime{\rm Improved~RPA}}}
\approx 3B.
$$

\chapter{Conclusions}
In this article we have initiated a detailed quantitative
study of the ``wee parton'' formulation of string theory
proposed in \Refer{thornweeparton,thornmosc}. Although
the quantitative relationship between the microscopic string
bit dynamics and the low energy stringy properties of
string theory is subsumed in merely  a renormalization of
the string tension, we believe that these details are
extremely important for understanding any processes which
are capable of probing the microscopic structure of the
theory. A gedanken experiment which could
in principle reveal this structure might
be the physics near the horizon of a black hole
as perceived by an observer far away\[thooftbh,susskindtu,susskindbh].
It is, of course, unlikely that the specific string
bit model we have chosen to analyze in this article is
the ultimate underlying dynamics of the bosonic string.
Our purpose, rather, is to pick {\it a} model and explore
its detailed dynamics using well-known methods in many
body physics. The model we have chosen, with
its delta function interaction potential, does nonetheless
have several
appealing features which recommend it, including locality
in the transverse space and (anomalous) scale invariance.

For pedagogical simplicity we have only tried to account
for the bosonic string in this article. But one reason
for subjecting the string bit dynamics for this
unrealistic model to such scrutiny is that the
superstring could be much more sensitive to the
microscopic details than is the bosonic string. In that
case we need to have a high degree of confidence in
the techniques used to analyze a superstring bit
model. We hope to devise such a model in the near future.

The numerical results of our study are encouraging.
Although the most na\"ive application of RPA to the
calculation of the rest tension is subjected
to huge (150\%) corrections in next order of perturbation
theory, we have shown that the bulk of these corrections
can be incorporated in a Hartree-Fock like improvement
of the zeroth order RPA. The corrections that remain are
estimated to be around 15\%. We did make justifiable
simplifications of the Hartree-Fock treatment, but there
is certainly room for improvement. It would be particularly
desirable to go beyond the lowest order approximation for
$\Sigma$. Also, in calculating $\Sigma^\prime$ we truncated
the Green function $G^\prime$ by dropping the ``free''
contribution which was shown to be relatively small in
perturbation theory. However, it should be a tractable
problem without this truncation. Thus, although the
simplifications we made were consistent with our
approximation, we expect that they can be avoided
with a little more effort.

$\underline{{\rm Acknowledgements.}}$ The older work reviewed in this
article and published in \Refer{thornweeparton}
benefited from valuable discussions with
Jeffrey Goldstone. I thank him again for his
direct contributions,
cited in the text, as well as for his insights
and advice on theoretical methods in many body
physics.  I also thank the Aspen Center for Physics where
the manuscript for this article was completed.

\appendix{}
In this appendix, we describe the evaluation of the various
matrix elements and integrals occurring in \eq{group1} and
\eq{group2}. The Green function occurring in group 1
is not singular in the integration range, so all $\epsilon$'s
may be set to zero. Then there is no distinction between
${\hat G}$ and $G$, so referring to \eq{green1alt} and
going to momentum representation, we may write
$${\hat G}_{-t-B}={1\over{\bf p}^2+B+t}+{\ket{t+B}\bra{t+B}\over
B\ln(1+t/B)},\(hat)$$
where we have introduced the state $\ket{t+B}$ defined by
$$\VEV{{\bf p}|t+B}\equiv\sqrt{B\over\pi}{1\over{\bf p}^2+B+t}.$$
Defining
$$\ket{B^\prime}\equiv {d\over dt}\ket{t+B}|_{t=0},$$
we can also write
$${\tilde G}_{-B}={1\over {\bf p}^2+B}+{1\over2B}\ket{B}\bra{B}
+\ket{B^\prime}\bra{B}+\ket{B}\bra{B^\prime}.\(tilde)$$

We first consider the contributions due to the first terms
in \eq{hat} and \eq{tilde}, the ``free'' Green functions.
The first four matrix elements in \eq{group1} turn out to equal each
other. Taking the first one, for example, we get
$$\eqalign{
&{B\over\pi}\int d{\bf p}{2ip^kp^m\over({\bf p}^2+B)^3}
{1\over({\bf p}^2+B+t)^2}i{\partial\over\partial p^j}
{p^m\over({\bf p}^2+B+t)({\bf p}^2+B)}=\cr
&\delta_{kj}B\int_B^\infty du
\left\{\left[{2B^2\over u^4}-{4B\over u^3}+{2\over u^2}\right]
{1\over(u+t)^3}
+\left[{2B^2\over u^5}-{3B\over u^4}+{1\over u^3}\right]{1\over(u+t)^2}
\right\}\cr}
$$
Let us define
$$I_{n,m}\equiv\int_1^\infty dv{v-1\over v^2}\int_1^\infty du
{1\over u^n(u+v)^m}. \(inm)$$
Needed numerical values for some of the $I_{n,m}$ are displayed
in the following table.

\bigskip
\begintable
\multispan{6}\tstrut\hfil $I_{n,m}$\hfil
          \crthick
n\\ m | 1 | 2 | 3 | 4 | 5 \cr
0\hfill |          |          |          | 0.007530 | 0.001851 \cr
1\hfill | 0.886294 | 0.090863 | 0.017005 | 0.004129 | 0.001163 \cr
2\hfill | 0.295431 | 0.045431 | 0.010202 | 0.002751 | 0.000832 \cr
3\hfill | 0.170431 | 0.029519 | 0.007168 | 0.002037 |          \cr
4\hfill | 0.118925 | 0.021702 | 0.005493 |          |          \cr
5\hfill | 0.091148 | 0.017114 |          |          |          \cr
6\hfill | 0.073836 |          |          |          |          \endtable
\bigskip

Then the ``free'' parts of the first four lines of \eq{group1}
evaluate to
$${-4\delta_{rs}\delta_{jk}\over B^2}
[2I_{4,3}-4I_{3,3}+2I_{2,3}+2I_{5,2}-3I_{4,2}+I_{3,2}]
\approx {-0.005436\delta_{rs}\delta_{jk}\over B^2}.
\(1234free1)$$
The 5th and 8th matrix elements are also equal, and their
free contributions to \eq{group1} are easily shown to give
$${-4\delta_{rs}\delta_{jk}\over B^2}
[I_{6,1}-2I_{5,1}+I_{4,1}]
\approx{-0.04186\delta_{rs}\delta_{jk}\over B^2}.\(58free1)
$$
The 6th and 7th matrix elements in group 1 are the most tedious
to evaluate, but the steps are straightforward, with the
result for their free contribution to \eq{group1}
$$\eqalign{
{-2\delta_{rs}\delta_{jk}\over B^2}
[2I_{2,5}-4I_{1,5}+2I_{0,5}
+4I_{3,4}-6I_{2,4}+2I_{1,4}+2I_{4,3}-&2I_{3,3}+I_{2,3}]
\approx\cr
&{-0.014932\delta_{rs}\delta_{jk}\over B^2}.\cr}
\(67free1)
$$
Finally the free contribution to \eq{group2} is the same
for each of the 4 matrix elements, leading to a total
contribution
$${-4\delta_{rs}\delta_{jk}\over B^2}
[-6I_{6,1}+10I_{5,1}-4I_{4,1}]
\approx{+0.028944\delta_{rs}\delta_{jk}\over B^2}.\(1234free2)
$$
We record here the total contribution from the ``free'' term
of the Green function to \eq{group1} and \eq{group2},
the sum of \eq{1234free1}, \eq{58free1}, \eq{67free1},
and \eq{1234free2},
$-0.033284/B^2$, which is approximately 10\% of the lowest
order (RPA) term.

Next we turn to the contribution from the second term of
\eq{hat} to the various terms in \eq{group1}. The evaluation
requires the matrix elements
$$\eqalign{
\bra{t+B}y^j{1\over{\bf p}^2+B+t}p^m\ket{B}=&{iB\delta_{jm}
\over\pi}\int d{\bf p}
{{\bf p}^2\over({\bf p}^2+B)({\bf p}^2+B+t)^3}\cr
=&{i\delta_{jm}\over B}\int_1^\infty du{u-1\over u(u+t/B)^3},\cr
\bra{t+B}p^m{1\over{\bf p}^2+B}y^j\ket{B}=&{-iB\delta_{jm}
\over\pi}\int d{\bf p}
{{\bf p}^2\over({\bf p}^2+B)^3({\bf p}^2+B+t)}\cr
=&{-i\delta_{jm}\over B}\int_1^\infty du{u-1\over u^3(u+t/B)},\cr}
$$
and their complex conjugates. The integrals here are elementary:
$$\eqalign{
\int_1^\infty du{u-1\over u(u+v)^3}=&{1\over v(v+1)}\left(
{1\over v}+\half-{1\over v}(1+{1\over v})\ln(1+v)\right)\cr
\int_1^\infty du{u-1\over u^3(u+v)}=&
{1\over v}
\left({1\over v}+\half-{1\over v}(1+{1\over v})\ln(1+v)\right).\cr}
$$
Using these results we find that the second term of \eq{hat}
inserted into \eq{group1} yields
$${4\delta_{rs}\delta_{jk}\over B^2}\int_1^\infty dt
{t-1\over t^4(1+t)\ln(1+t)}
\left({1\over t}+\half-{1\over t}(1+{1\over t})\ln(1+t)\right)^2
\approx{0.005795\delta_{rs}\delta_{jk}\over B^2}
\(1234pole1)
$$
for the first 4 lines of \eq{group1}. For the 5th and 8th
lines of \eq{group1} the corresponding contribution is
$${-2\delta_{rs}\delta_{jk}\over B^2}\int_1^\infty dt
{t-1\over t^4\ln(1+t)}
\left({1\over t}+\half-{1\over t}(1+{1\over t})\ln(1+t)\right)^2
\approx{-0.009937\delta_{rs}\delta_{jk}\over B^2}
\(58pole1)
$$
and for the 6th and 7th lines,
$$\eqalign{
{-2\delta_{rs}\delta_{jk}\over B^2}\int_1^\infty dt
{t-1\over t^4(1+t)^2\ln(1+t)}
&\left({1\over t}+\half-{1\over t}\left(1+{1\over t}\right)
\ln(1+t)\right)^2
\approx\cr
&\qquad\qquad\qquad\qquad\qquad
{-0.000967\delta_{rs}\delta_{jk}\over B^2}.\cr
}
\(67pole1)
$$
The numerical values for these integrals were obtained by
a simple integration routine.
For reference purposes we also quote the values of the terms
in \eq{1diagramred} that have been cancelled, so that the total
value of the first diagram in \Fig{corrections} can be worked
out. This works out to a total of $-\delta_{rs}\delta_{kj}/(18B^2)$.
Noting that \eq{1234pole1} almost cancels \eq{1234free1}, we
see that this last bit is the dominant contribution to the
diagram, about 16\% of the RPA contribution.

The contribution from the last three terms of \eq{tilde} to
\eq{group2} requires the matrix elements
$$\eqalign{
\bra{B}y^j{1\over{\bf p}^2+B}y^k\ket{B}=&{\delta_{jk}\over 6B^2}\cr
\bra{B^\prime}y^j{1\over{\bf p}^2+B}y^k\ket{B}=&
-{\delta_{jk}\over 5B^2}\cr
\bra{B}p^m{1\over{\bf p}^2+B+t}p^m\ket{B}=&
\int_1^\infty du{u-1\over u^2(u+t/B)}\cr
\bra{B^\prime}p^m{1\over{\bf p}^2+B+t}p^m\ket{B}=&
-{1\over B}\int_1^\infty du{u-1\over u^3(u+t/B)}.\cr}
$$
Using these results, we then obtain for the remainder of
the evaluation of \eq{group2}
$$\eqalign{
{-4\delta_{rs}\delta_{jk}\over B^2}&\left[({1\over12}-{1\over5})
(I_{1,1}-I_{2,1})-{1\over6}(I_{2,1}-I_{3,1})\right]
=\cr
&\qquad\qquad\qquad{\delta_{rs}\delta_{jk}\over B^2}
\left[{7\over15}(I_{1,1}-I_{2,1})+{2\over3}(I_{2,1}-I_{3,1})\right]
\approx{0.359069\over B^2}\cr}
\(1234pole2)
$$
Finally, we evaluate the double pole contributions
\eq{2diagramdoub} and \eq{2diagramdoub}, which requires
the additional matrix elements
$$\eqalign{
\bra{B}y^j{1\over{(\bf p}^2+B)^2}y^k\ket{B}
=&{\delta_{jk}\over 10B^3}\cr
\bra{B}p^m{1\over({\bf p}^2+B+t)^2}p^m\ket{B}=&
{1\over B}\int_1^\infty du{u-1\over u^2(u+t/B)^2}.\cr}
$$
Then the total double pole contribution evaluates to
$$\eqalign{
{\delta_{rs}\delta_{jk}\over B^2}\left[{1\over5}
(I_{1,1}-I_{2,1})+{1\over3}(I_{1,2}-I_{2,2})\right]
\approx&{0.133316\over B^2}\cr}
\(1234pole2)
$$

The grand total of all contributions is
$${\rm TOTAL}\approx{0.453992\delta_{rs}\delta_{kj}\over B^2},$$
which is $-136\%$ of the RPA result. Clearly this cannot
be regarded as a small correction!

\vskip.3in
\titlestyle{REFERENCES}
\reflist{}
\bye